\documentclass[fleqn,usenatbib]{mnras}

\usepackage{newtxtext,newtxmath}

\usepackage{mathrsfs}
\usepackage[T1]{fontenc}
\usepackage{ae,aecompl}
\usepackage{booktabs}

\usepackage{graphicx}	% Including figure files
\usepackage{amsmath}	% Advanced maths commands
\usepackage{amssymb}	% Extra maths symbols
\usepackage{float}

\usepackage{empheq}

\title[Energy functions of CHIME FRBs]{Energy functions of fast radio bursts derived from the first CHIME/FRB catalogue}

\author[T. Hashimoto et al.]{
Tetsuya Hashimoto,$^{1}$\thanks{E-mail: tetsuya@phys.nchu.edu.tw}
Tomotsugu Goto,$^{2}$
Bo Han Chen,$^{2}$
Simon C.-C. Ho,$^{2}$
\newauthor
Tiger Y.-Y. Hsiao,$^{2}$
Yi Hang Valerie Wong,$^{2}$
Alvina Y. L. On,$^{2,3}$
Seong Jin Kim,$^{2}$
\newauthor
Ece Kilerci-Eser,$^{4}$
Kai-Chun Huang,$^{1}$
Daryl Joe D. Santos, $^{5}$
and Shotaro Yamasaki$^{1,6}$
\\
$^{1}$Department of Physics, National Chung Hsing University, No. 145, Xingda Rd., South Dist., Taichung, 40227, Taiwan (R.O.C.)\\
$^{2}$Institute of Astronomy, National Tsing Hua University, 101, Section 2. Kuang-Fu Road, Hsinchu, 30013, Taiwan (R.O.C.)\\
$^{3}$Mullard Space Science Laboratory, University College London, Holmbury St Mary, Surrey RH5 6NT, UK\\
$^{4}$Sabanc{\i} University, Faculty of Engineering and Natural Sciences, 34956, Istanbul, Turkey\\
$^{5}$Max Planck Institute for Extraterrestrial Physics, Gie{\ss}enbachstra{\ss}e 1, 85748 Garching, Germany\\
$^{6}$Racah Institute of Physics, The Hebrew University of Jerusalem, Jerusalem 91904, Israel
}

\date{Accepted 2022 January 3. Received 2022 January 3 in original form 2021 October 18}

\pubyear{2022}

\begin{document}
\label{firstpage}
\pagerange{\pageref{firstpage}--\pageref{lastpage}}
\maketitle

\begin{abstract} %250 currently 250 limit
Fast radio bursts (FRBs) are mysterious millisecond pulses in radio, most of which originate from distant galaxies. 
Revealing the origin of FRBs is becoming central in astronomy. 
The redshift evolution of the FRB energy function, i.e., the number density of FRB sources as a function of energy, provides important implications for the FRB progenitors. 
Here we show the energy functions of FRBs selected from the recently released Canadian Hydrogen Intensity Mapping Experiment (CHIME) catalogue using the $V_{\rm max}$ method. 
The $V_{\rm max}$ method allows us to measure the redshift evolution of the energy functions as it is without any prior assumption on the evolution. 
We use a homogeneous sample of 164 non-repeating FRB sources, which are about one order of magnitude larger than previously investigated samples.
The energy functions of non-repeating FRBs show Schechter function-like shapes at $z\lesssim1$. 
The energy functions and volumetric rates of non-repeating FRBs decrease towards higher redshifts similar to the cosmic stellar-mass density evolution: 
there is no significant difference between the non-repeating FRB rate and cosmic stellar-mass density evolution with a 1\% significance threshold, whereas the cosmic star-formation rate scenario is rejected with a more than 99\% confidence level.
Our results indicate that the event rate of non-repeating FRBs is likely controlled by old populations rather than young populations which are traced by the cosmic star-formation rate density. 
This suggests old populations such as old neutron stars and black holes as more likely progenitors of non-repeating FRBs.
%Based on well-explored FRBs with robust measurements and known selection functions, these results are solid. %Extending the observational parameter space towards possible missing populations with high scattering and low fluence is essential to fully understand the FRB population.
%For robust measurements, FRBs with large scattering or low fluence are not used in this work. Fulling sampling FRBs including these would be important in the future.
\end{abstract}

\begin{keywords}
radio continuum: transients -- stars: magnetars -- stars: magnetic field -- stars: neutron -- (stars:) binaries: general -- stars: luminosity function, mass function
\end{keywords}

%%%%%%%%%%%%%%%%%%%%%%%%%%%%%%%%%%%%%%%%%%%%%%%%%%

%%%%%%%%%%%%%%%%% BODY OF PAPER %%%%%%%%%%%%%%%%%%

\section{Introduction}
\label{introduction}
Since the first discovery of a fast radio burst \citep[FRB;][]{Lorimer2007}, more than 600 FRBs have been detected \citep[e.g.][]{Petroff2016,CHIMEcat2021}.
Numerous FRB theories have been proposed \citep[e.g.][]{Platts2019} for non-repeating and repeating FRBs, where non-repeating and repeating FRBs are observationally defined as a one-off burst and multiple bursts detected from each FRB source, respectively.
Despite the intensive FRB observations and their modelling so far, the origin of most FRBs is still unknown.

Accurate localisation of FRB sources is one of the most straightforward and powerful approaches to the identification of the FRB progenitor.
A case of direct identification of an FRB progenitor is repeating FRB 200428 which was localised at the position of a Galactic magnetar, SGR 1935+2154 \citep[e.g.][]{Scholz2020,Bochenek2020,Kirsten2020}.  
However, FRB 200428 is about 30 times less energetic than the faintest population of typical extragalactic FRBs \citep{Bochenek2020,Marcote2020}.
%Therefore, whether FRB 200428 is an archetype of extragalactic FRBs or not is still in debate.
Therefore, whether progenitors of extragalactic FRBs are also magnetars or not is still in debate.

For extragalactic FRBs, one of the localised ones is FRB 180916.J0158+65 \citep{Marcote2020}.
The repeating FRB source of FRB 180916.J0158+65 is located at the vicinity of a star-forming region in a spiral galaxy at $z=0.0337$ \citep{Marcote2020}.
Another case of repeating FRB source is 20201124A \citep{Piro2021}, which is also localised at a star-forming region in a nearby galaxy at $z=0.0978$ \citep{Piro2021}.
These observations suggest that star formation and thus young populations may be related to the FRB progenitors.
On the contrary, another repeating FRB source, 2020120E, is localised at the position of a globular cluster in M81 \citep{Bhardwaj2021,Kirsten2021}, suggesting old populations as the progenitor of this FRB source. 
Recently, \citet{Xu2021} reported that the actively repeating FRB source, 20201124A, is located at an inter-arm region of a barred-spiral galaxy at redshift $z=0.09795$, suggesting an environment not directly expected for young populations.
Even such well-localised cases seem to show diverse environmental properties of FRB progenitors.
Such diverse environmental properties of FRBs are also reported by observations of FRB host galaxies \citep[e.g.][]{Bhandari2020,Bhandari2021,Lorimer2021}.
However, the number of localised FRBs is currently only $\sim20$ \citep[e.g.][]{Bhandari2021}, hampering precise statistics.
This is because FRB localisation requires a high spatial resolution in radio and multi-wavelength follow-up observations, which are expensive and time-consuming in general.

An alternative approach to probing the FRB origin is to investigate the \lq FRB population\rq.
The number density of FRB sources can be compared with that of possible progenitors to constrain the FRB origin \citep[e.g.][]{Ravi2019,Luo2020,Hashimoto2020a}.
The redshift evolution of the luminosity or energy functions of FRBs, i.e. number density of FRB sources as a function of luminosity or energy, is one of the useful tools to constrain the FRB progenitors \citep[e.g.][]{Hashimoto2020c,Arcus2021,James2022}.
If FRB progenitors are young, i.e. produced via star formation, the number density of FRBs should increase towards higher redshifts up to $z\sim2$ because the cosmic star-formation rate density increases towards higher redshifts \citep[e.g.][]{Madau2014,Madau2017}.
In contrast, the FRB number density may decrease towards higher redshifts if old populations such as old neutron stars and black holes are predominant as FRB progenitors. 
Such statistical analyses of the FRB population allow the detected FRBs to be fully utilised regardless of localisation.

\citet{James2022} constructed an FRB population model to fit with the observed distribution of dispersion measures using Australian Square Kilometre Array Pathfinder (ASKAP) and Parkes FRBs. 
They reported that the FRB luminosity function evolves similarly to (or faster than) the cosmic star-formation rate density, assuming that the evolution is scaled with the cosmic star-formation rate density with a free power-law parameter.
\citet{Arcus2021} performed a similar approach to that of \citet{James2022} to show that either the cosmic star-formation rate density evolution or no evolution can explain the observed distribution of dispersion measures of ASKAP and Parkes FRBs.
\citet{Zhangetal2021} tested the observed Parkes and ASKAP samples on redshift distribution models tracking the two extremes of evolving redshift distribution models (star-formation rate history and compact binary merger history).
They found that the limited data sample was consistent with both of those models.
\citet{Hashimoto2020c} used the $V_{\rm max}$ method \citep[][see also Section \ref{vmax} for details]{Schmidt1968} to directly measure the number density of Parkes non-repeating FRBs without any prior assumption on the redshift evolution.
They found that the number density of non-repeating FRB sources does not show any significant redshift evolution up to $z\sim2$, which is consistent with the cosmic stellar-mass density evolution rather than the cosmic star-formation rate density.

These FRB population analyses mentioned above have shown diverse results.
This could be due to (i) that \citet{James2022} and \citet{Arcus2021} do not test old population models or (ii) the small number of FRB samples which are less than 100 in these works \citep{Hashimoto2020c,Arcus2021,Zhangetal2021,James2022}.

\citet{CHIMEcat2021} released the new Canadian Hydrogen Intensity Mapping Experiment (CHIME) FRB catalogue which includes 536 FRB events detected over an effective survey duration of 214.8 days, of which 474 are unique non-repeating FRB sources and 18 are repeating FRB sources.
The new CHIME FRBs allow a much better statistical analysis of the FRB population with about one order of magnitude larger homogeneous sample than that in previous works.
\citet{Zhang2021} tested the new CHIME sample against the star-formation rate density, cosmic stellar-mass density, and delayed models.
They found that the models including significant delays of FRBs ($\gtrsim10$ Gyr) with respect to star formation better describe the observed distributions of fluences, energies, and dispersion measures than others, suggesting the old population as the origin of FRBs.

In this work, we present energy functions and volumetric rates of the new CHIME FRBs as a function of redshift to constrain the FRB progenitor without any prior assumption on the redshift evolution.
Throughout this paper, we use the terminology of \lq rest-frame\rq\ to refer to a frame of an FRB source.
The {\it Planck15} cosmology \citep{Planck2016} is adopted as a fiducial model, i.e., $\Lambda$ cold dark matter cosmology with ($\Omega_{m}$,$\Omega_{\Lambda}$,$\Omega_{b}$,$h$)=(0.307, 0.693, 0.0486, 0.677), unless otherwise mentioned.

The structure of the paper is as follows: we describe how selection functions, energy functions, and volumetric rates are derived in Section \ref{analysis}.
In Section \ref{results}, we present the energy functions and volumetric rates of non-repeating and repeating FRB sources along with their redshift evolution.
The indications of our results on non-repeating FRBs and their possible origins are discussed in Section \ref{discussion} followed by conclusions in Section \ref{conclusion}.

\section{Data analysis}
\label{analysis}
\subsection{Selection functions}
\label{selection_function}
\citet{CHIMEcat2021} tested the CHIME detection algorithm by injecting 84,697 simulated FRB signals of which 39,638 were detected.
They found that significant fractions of the injected mock FRBs with long scattering times or low fluences are missed by the CHIME detection algorithm. 
In addition, moderate fractions of FRBs with small or large observed dispersion measures (DM$_{\rm obs}$) and FRBs with long intrinsic durations ($w_{\rm int}$) are also missed by the algorithm.
The observed data distribution in the parameter space of spectral index and running is reasonably reproduced by the simulated FRB detection \citep{CHIMEcat2021}.
Therefore, following \citet{CHIMEcat2021}, we consider selection functions of dispersion measure, scattering time, intrinsic duration, and fluence in this work.
These selection functions have to be known to correctly calculate the FRB energy functions and the FRB number densities.
In this work, we use the signal-to-noise (SNR) cut at SNR $=10$ to maintain a meaningful number of repeating FRBs in our sample.
This SNR cut is slightly lower than SNR $=12$ which is used in \citet{CHIMEcat2021}.
However, \citet{CHIMEcat2021} verified that their conclusions hold when all catalogue events are included regardless of SNR.

\subsubsection{Sample for selection functions}
\label{sample_SF}
The selection functions are provided only for the case of SNR cut $=12$ in \citet{CHIMEcat2021}.
Therefore, we empirically derive the selection functions for the case of SNR cut $=10$ in this work.
We first define the sample to derive the selection functions.
The selected sample satisfies all of the following criteria.
\begin{itemize}
\item {\it bonsai\_snr} $>10$
\item DM$_{\rm obs} >1.5 \times$max(DM$_{\rm NE2001}$, DN$_{\rm YMW16}$)
\item not detected in far side-lobes
\item $\log \tau_{\rm scat} < 1.0$ (ms)
\item {\it excluded\_flag} $=0$
\item the first detected burst if the FRB source is a repeater
\item fluence $>0$ (Jy ms),
\end{itemize}
where $\tau_{\rm scat}$ is the scattering time \citep[see][for details]{CHIMEcat2021}.
These criteria are the same as that used in \citet{CHIMEcat2021} except for the SNR cut.
When the only upper limit is available for $\tau_{\rm scat}$, we utilise its 1 $\sigma$ upper limit as $\tau_{\rm scat}$, following \citet{CHIMEcat2021}.
After applying these selection criteria to the new CHIME catalogue, the selected sample includes 348 non-repeating and 13 repeating FRB sources.
We note that \citet{CHIMEcat2021} used 265 FRB sources to derive selection functions with SNR cut $=12$.

To derive the selection functions, we follow Equation 6 in \citet{CHIMEcat2021}, which describes the relationship between the observed and intrinsic data distributions for a certain parameter such as dispersion measure:
\begin{equation}
\label{eq:selection}
P({\rm DM})=P_{\rm obs}({\rm DM}) \times s({\rm DM})^{-1},
\end{equation}
where $P({\rm DM})$ and $P_{\rm obs}({\rm DM})$ are the intrinsic and observed distributions of dispersion measure, respectively.
$s({\rm DM})$ is the selection function as a function of the dispersion measure.
The intrinsic data distributions for the case of SNR cut $=10$ are provided in \citet{CHIMEcat2021}:
\begin{equation}
\label{Pxi}
P(\xi)=\frac{1}{\sigma(\xi/m)\sqrt{2\pi}} \exp\left[ -\frac{\ln^{2}(\xi/m)}{2\sigma^{2}}\right],
\end{equation}
where $\xi$ could either be dispersion measure, scattering time, and intrinsic duration.
The scale $m$ is 512 (pc cm$^{-3}$) for the dispersion measure, 2.04 (ms) for the scattering time, and 1.19 (ms) for the intrinsic duration.
The shape $\sigma$ is 0.68, $1.57$, and 0.99, respectively.
The scale $m$ and shape $\sigma$ represent characteristic values of the parameters and their dispersions, respectively, describing the intrinsic data distributions of the CHIME FRBs via Eq. \ref{Pxi} \citep{CHIMEcat2021}.
The intrinsic fluence distribution is
\begin{equation}
\label{PFnu}
P(F_{\nu}) \propto -\alpha (F_{\nu}/F_{\nu,0})^{\alpha-1},
\end{equation}
where $\alpha$ is the power-law index and $F_{\nu,0}$ is an arbitrary pivot fluence.
We use $\alpha=-1.41$ in this work.

Using Eq. \ref{eq:selection}, $s({\rm DM})$ can be empirically derived by dividing $P_{\rm obs}({\rm DM})$ by $P({\rm DM})$.
The selection functions for the other parameters including scattering time, intrinsic duration, and fluence are also derived in the same manner as Eq. \ref{eq:selection}.

\subsubsection{Derived selection functions}
\label{derived_SF}
\begin{figure*}
    \includegraphics[width=0.8 \columnwidth]{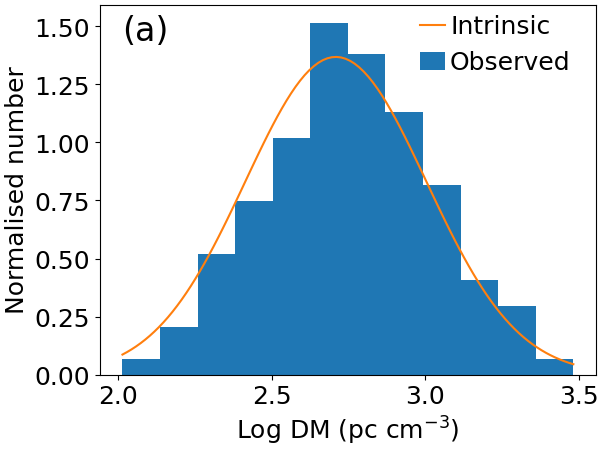}
    \includegraphics[width=0.8 \columnwidth]{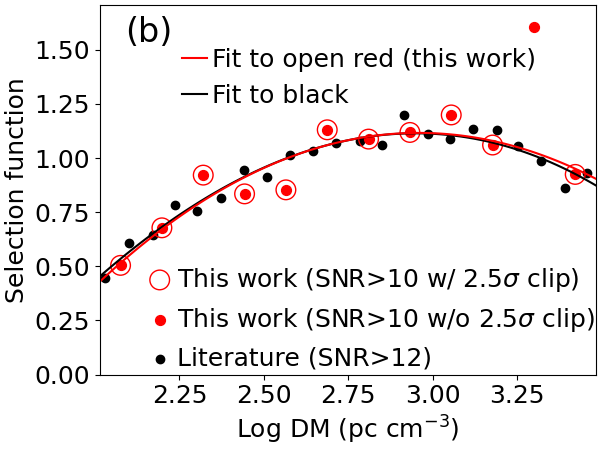}
    \includegraphics[width=0.8 \columnwidth]{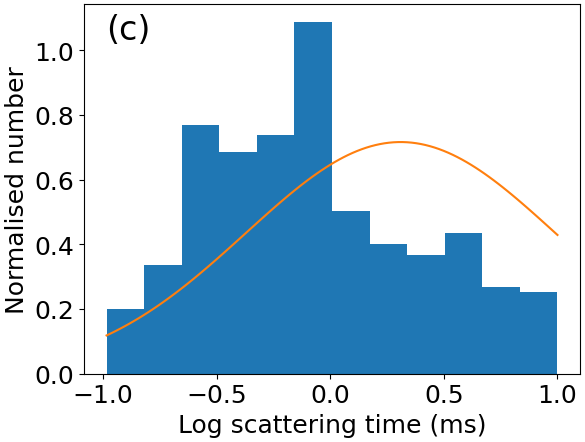}
    \includegraphics[width=0.8 \columnwidth]{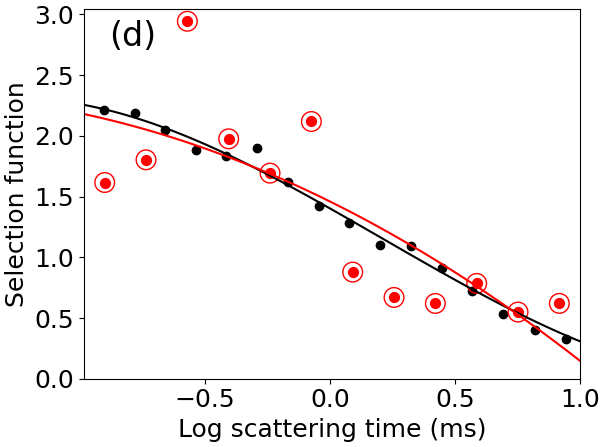}
    \includegraphics[width=0.8 \columnwidth]{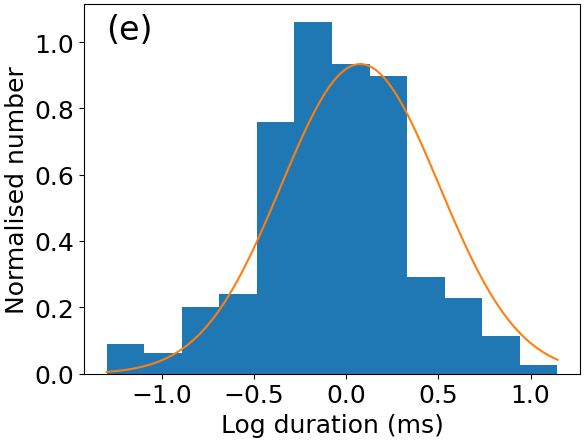}
    \includegraphics[width=0.8 \columnwidth]{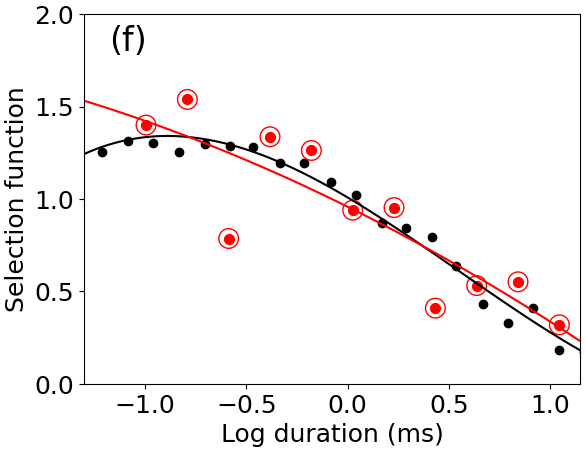}
    \includegraphics[width=0.8 \columnwidth]{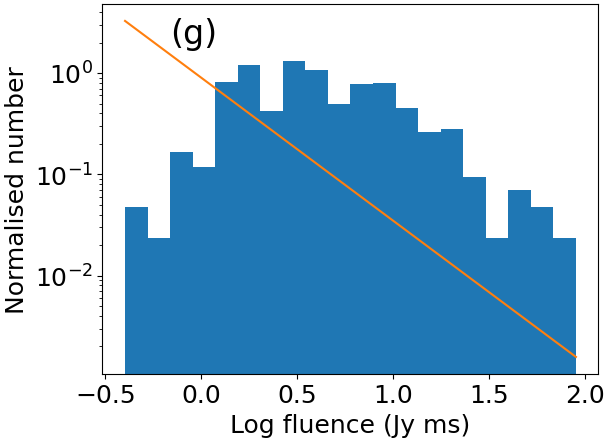}
    \includegraphics[width=0.8 \columnwidth]{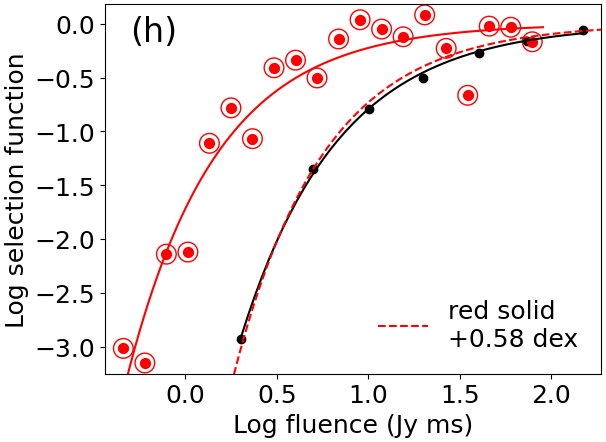}
    \caption{
    From top to bottom panels, dispersion measure, scattering time, intrinsic duration, and fluences of CHIME FRBs are shown.
    The observed and intrinsic data distributions (filled histograms and orange solid line, respectively) are shown in the left panels while the derived selection functions are presented in the right panels.
    The intrinsic data distribution refers to $P(\xi)$ or $P(F_{\nu})$ in Eqs. \ref{Pxi} and \ref{PFnu}.
    The presented sample is selected based on the criteria described in Section \ref{sample_SF}. 
    The intrinsic distributions for the case of SNR cut $=10$ are adopted from \citet{CHIMEcat2021}.
    Both observed and intrinsic data distributions are normalised.
    In the right panels, the red dots indicate selection functions derived in this work.
    The selection functions after 2.5 $\sigma$ clipping are marked by red open circles.
    The red solid lines indicate polynomial fittings to the red open circles.
    The selection functions for the case of SNR cut $=12$ \citep{CHIMEcat2021} and its polynomial fittings are shown by black dots and black solid lines, respectively.
    In the panel (h), the red solid line indicates observed fluences, whereas the black solid line shows the intrinsic fluences of injected mock FRBs \citep{CHIMEcat2021}.
    The red solid line is horizontally shifted by $+0.58$ dex (red dashed line) for a comparison with the SNR cut $=12$ case (black solid line).
    }
    \label{fig1}
\end{figure*}

The derived selection functions are summarised in Fig. \ref{fig1}.
In Fig. \ref{fig1}, the observed and intrinsic data distributions are presented in the left panels while the derived selection functions for the case of SNR cut $=10$ are shown by red dots in the right panels.
We fit polynomial functions to the derived selection functions of the dispersion measure, scattering time, and intrinsic duration with a 2.5$\sigma$ clip.
For the selection function (Eq. \ref{eq:selection}) of fluence, the following functional shape is adopted
\begin{equation}
\log s({\rm SNR}>10|F_{\nu})=a(1.0-\exp(-b\log F_{\nu}))-a,
\end{equation}
where $a$ and $b$ are fitting parameters such that the function converges to 0 assuming that the FRB detection is 100\% complete at high fluences, i.e., $\log s({\rm SNR}>10|F_{\nu})=0$.
The data points used for these fitting procedures are highlighted by red open circles in the right panels.

In Fig. \ref{fig1}, the selection functions for the SNR cut $=10$ (red solid lines) are almost the same as that of SNR cut $=12$ presented in \citet{CHIMEcat2021} (black solid lines) except for the fluence selection function shown at the panel (h) of Fig. \ref{fig1}.
The observed fluences of CHIME observations are uncertain because the localisation of each FRB is not accurate except for FRB 121102 at $z=0.19273$ \citep{Tendulkar2017} and FRB 180916.J0158+65 at $z=0.0337$ \citep{Marcote2020} which were localised by other telescopes.
The observed fluences are lower limits because the telescope sensitivity at the centre of the field of view is assumed \citep{CHIMEcat2021}.
Therefore, there should be an offset between the observed fluences and true fluences on average.
In panel (h) in Fig. \ref{fig1}, we use the observed fluences for the case of SNR cut $=10$ (red solid line), whereas the case of SNR cut $=12$ (black solid line) indicates true fluences of injected mock FRBs \citep{CHIMEcat2021}.
Indeed, we find an offset between the two, which is $\sim0.58$ dex in the logarithmic scale.
In this paper, we do not correct for this offset because it does not affect the FRB number density integrated over the energy (see Section \ref{derived_FRB_rate} for details) which is the main focus of this paper.
Except for this offset, the fluence selection function for the case of SNR cut $=10$ shows almost the same shape as that of SNR cut $=12$.
The best-fit functions are 
\begin{equation}
\label{selectionDM}
s({\rm SNR}>10|{\rm DM})=-0.7707(\log{\rm DM})^{2}+4.5601(\log{\rm DM})-5.6291
\end{equation}
\begin{equation}
s({\rm SNR}>10|\tau_{\rm scat})=-0.2922(\log\tau_{\rm scat})^{2}-1.0196(\log\tau_{\rm scat})+1.4592
\end{equation}
\begin{equation}
s({\rm SNR}>10|w_{\rm int})=-0.0785(\log w_{\rm int})^{2}-0.5435(\log w_{\rm int})+0.9574
\end{equation}
\begin{equation}
\label{selectionfnu}
\log s({\rm SNR}>10|F_{\nu})=1.7173(1.0-\exp(-2.0348\log F_{\nu}))-1.7173.
\end{equation}
These selection functions are utilised in calculating the number densities of FRBs in Section \ref{calc_density}.
Following \citet{CHIMEcat2021}, 1 $\sigma$ upper limits are adopted as $\tau_{\rm scat}$ and $w_{\rm int}$ in this section when only upper limits on them are available.
This treatment of the upper limits is also the case when the selection functions are applied to the number density calculation described in Section \ref{calc_density}.

\subsection{Calculations of physical parameters}
\label{physpara}
\subsubsection{Redshift}
\label{calc_redshift}
We follow the same manner as that of \citet{Hashimoto2020c} to derive the redshift of each FRB using the observed dispersion measure.
The derived redshifts are consistent with the spectroscopic redshifts of the host galaxies within the uncertainties \citep{Hashimoto2020c}.
Here, we briefly describe how redshift and its uncertainty are calculated for each FRB.
The observed dispersion measure (DM$_{\rm obs}$) is composed of four components including interstellar medium in the Milky Way (DM$_{\rm MW}$), extended hot gas associated with the dark matter halo hosting the Milky Way (DM$_{\rm halo}$), intergalactic medium (DM$_{\rm IGM}$), and the FRB host galaxy (DM$_{\rm host}$):
\begin{equation}
\label{eq:DM}
{\rm DM}_{\rm obs}={\rm DM_{\rm MW}}+{\rm DM}_{\rm halo}+{\rm DM}_{\rm IGM}+{\rm DM}_{\rm host}.
\end{equation}
We adopt DM$_{\rm MW}$ modelled by \citet{Yao2017}, DM$_{\rm halo}=65$ pc cm$^{-3}$ \citep{Prochaska2019}, and DM$_{\rm host}=50.0/(1+z)$ pc cm$^{-3}$ following \citet{Macquart2020}. 
The DM$_{\rm IGM}$ averaged over the line-of-sight fluctuation is described as a function of redshift with some assumptions on the cosmological parameters \citep[e.g. Equation 2 in][]{Zhou2014}.
Therefore, Eq. \ref{eq:DM} is expressed as a function of redshift. 
The solution provides each FRB with a redshift.
Under these assumptions, the error of DM$_{\rm obs}$ ($\delta$DM$_{\rm obs}$) and the line-of-sight fluctuation of DM$_{\rm IGM}$ ($\sigma_{\rm DM_{\rm obs}}$) contribute to the redshift uncertainty via error propagation in Eq. \ref{eq:DM}.
To estimate the redshift uncertainty of each FRB, we performed Monte Carlo (MC) simulations.
In each simulation, randomised errors are added to DM$_{\rm obs}$ and DM$_{\rm IGM}$. 
The randomised errors follow Gaussian probability distributions with standard deviations of $\delta$DM$_{\rm obs}$ and $\sigma_{\rm DM_{\rm obs}}$.
We conservatively assume the highest $\sigma_{\rm DM_{\rm obs}}$ estimated from cosmological simulations of structure formation \citep{Zhu2018}.
Since $\sigma_{\rm DM_{\rm obs}}$ is estimated as a function of redshift up to $z=2$ \citep{Zhu2018}, we linearly extrapolate $\sigma_{\rm DM_{\rm obs}}$ towards higher redshifts \citep[see][for details]{Hashimoto2020c}. 
For each FRB, the simulations were repeated 10,000 times to estimate the probability distribution of the redshift.
The 50 percentile of the distribution (median) is adopted as the redshift of each FRB.
We use the 84 and 16 percentiles as $\pm1 \sigma$ uncertainty.

\begin{figure}
    \includegraphics[width=\columnwidth]{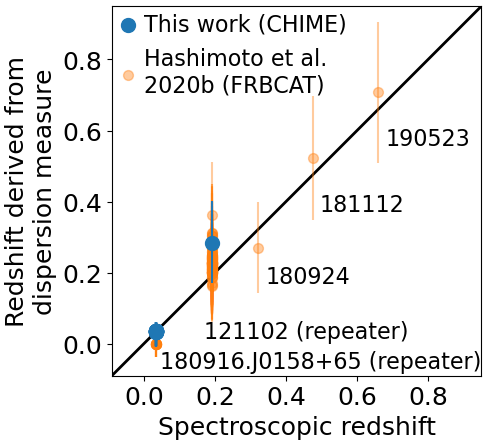}
    \caption{
    The redshift derived from the dispersion measure as a function of spectroscopic redshift for CHIME FRB sources (blue dots).
    The vertical errors are calculated by the error propagation of observational uncertainties of dispersion measures and line-of-sight fluctuation of the intergalactic dispersion measures (see Section \ref{calc_redshift} for details).
    Orange dots indicate redshifts of other FRB host galaxies in the FRB Catalogue project \citep[FRBCAT;][]{Petroff2016} derived by \citet{Hashimoto2020c} with the same assumptions on DM$_{\rm halo}$ and DM$_{\rm host}$ as those in Section \ref{calc_redshift}.
    }
    \label{fig2}
\end{figure}

In the first CHIME/FRB catalogue, there are two FRB sources with measurements of spectroscopic redshifts, i.e., FRB 121102 at $z=0.19273$ \citep{Tendulkar2017} and FRB 180916.J0158+65 at $z=0.0337$ \citep{Marcote2020}.
For these FRB sources, we utilise their spectroscopic redshift instead of the redshifts derived from Eq. \ref{eq:DM} in the following analyses. 
%\textcolor{cyan}{[SY: I was wondering why only CHIME FRBs are shown in Fig. \ref{fig2}. The methodology for determining redshift is generic and independent of FRB sample selection. Showing all the localised FRBs available (not only CHIME sample) would better demonstrate the power/limitation of the method used here. ]}
The spectroscopic redshifts are shown in Fig. \ref{fig2} compared with the redshifts derived from Eq. \ref{eq:DM}.
We confirmed that the redshifts derived from the dispersion measures are consistent with the spectroscopic redshift within the uncertainties. 
%\textcolor{cyan}{[SY: Another minor comment: the excess in the derived redshift for FRB 121102 is due to the underestimation of DM$_{\rm host}$. Observations suggest that DM$_{\rm host}$ for this source is significantly larger than the value assumed here. This demonstrates that the potential uncertainty in DM$_{\rm host}$, which is neglected here, could be an additional error in the derived redshift. Although a higher value of DM$_{\rm host}$ is tried in the Appendix, DM$_{\rm halo}$ is also reduced simultaneously (as a result DM$_{\rm host}+$DM$_{\rm halo}$ does not vary significantly), which may obscure the situation.]}
This point is also presented in \citet{Hashimoto2020c} with more spectroscopic samples using other localised FRBs (see also Fig. \ref{fig2}) in the FRB Catalogue project \citep[FRBCAT;][]{Petroff2016}.

\subsubsection{Energy integrated over the rest-frame 400 MHz width}
\citet{Hashimoto2020a} and \citet{Hashimoto2020c} utilise the time-integrated luminosity in units of erg Hz$^{-1}$, which is calculated from the observed fluence.
In this work, we use the energy in units of ergs, i.e., fluence integrated over the frequency, as an indicator of the brightness of FRBs based on the following reasons.
The first reason is that some FRBs show complicated sub-structures in their light curves.
Such shapes of light-curve and peak flux density would highly depend on the time resolutions of instruments. 
The fluence is less affected by the finite time resolution of instruments \citep[e.g.][]{Macquart2018}, which allows us to mitigate systematic differences when comparing with FRBs detected with other telescopes \citep[e.g.][]{Hashimoto2020a,Hashimoto2020c}.
The second reason is that FRBs detected with CHIME also show complicated spectral shapes \citep{Pleunis2021}.
\citet{Pleunis2021} presented the diverse spectral shapes: non-repeating FRBs tend to show broad-band power-law like shapes whereas repeating FRBs tend to show narrow-band Gaussian-like spectral shapes.
The $k$-correction for such diverse spectral shapes would be highly uncertain since complicated extrapolations of the spectral shapes are necessary.
To minimise such uncertainty, we integrate the fluence over the frequency to calculate observed energy ($E_{\rm obs}$) for each FRB. 
This frequency integration is described as $E_{\rm obs}=$ fluence $\times\left(\frac{400\times10^{6}}{{\rm Hz}}\right)$ because the fluence in the first CHIME/FRB catalogue is the band-averaged value over the CHIME frequency width of 400 MHz \citep{CHIMEcat2021}.

The observed frequency width of 400 MHz corresponds to different frequency widths at the rest-frame depending on the redshifts of FRBs.
For a fair comparison at different redshifts, we use the integration over 400 MHz widths at the rest-frame.
We define the integration width in the observer-frame, $\Delta\nu_{\rm obs,itg}$, which corresponds to the 400 MHz at the rest-frame, i.e., $\Delta\nu_{\rm obs,itg}=400/(1+z)$ MHz.
The observed energy integrated over the rest-frame 400 MHz width ($E_{\rm obs,400}$) is approximated as follows:
\begin{equation*}
E_{\rm obs,400}=
\end{equation*}
\vspace{-20pt}
\begin{empheq}[left={\empheqlbrace}]{alignat=1}
\label{eq:Eobs400_1}
& F_{\nu}\left(\frac{400\times10^{6}}{{\rm Hz}}\right) ~~~~~~~~~~~~~~~~~~~~~~~~~(\Delta\nu_{\rm obs,itg}\geq \Delta\nu_{\rm obs,FRB})\\
\label{eq:Eobs400_2}
& F_{\nu} \left(\frac{400\times10^{6}}{{\rm Hz}}\right)\left(\frac{\Delta\nu_{\rm obs,itg}}{\Delta\nu_{\rm obs,FRB}}\right) ~~~~(\Delta\nu_{\rm obs,itg}<\Delta\nu_{\rm obs,FRB}),
\end{empheq}
where $F_{\nu}$ is the observed fluence and $\Delta\nu_{\rm obs,FRB}$ is the frequency width in which FRB is detected.
For each FRB without multiple sub-bursts, $\Delta\nu_{\rm obs,FRB}$ is calculated by {\it high\_freq} $-$ {\it low\_freq} in the first CHIME/FRB catalogue, where {\it high\_freq} ({\it low\_freq}) is the highest (lowest) frequency band of detection at a full-width-tenth-maximum. 
For each FRB with multiple sub-bursts, the maximum (minimum) value of {\it high\_freq} ({\it low\_freq}) is adopted to calculate $\Delta\nu_{\rm obs,FRB}$ because there is no frequency gap between the sub-bursts in the catalogue.
The ratio, $\Delta\nu_{\rm obs,itg}/\Delta\nu_{\rm obs,FRB}$, approximately takes the overflowed energy out of the rest-frame 400 MHz width into account.

Following \citet{Macquart2018b}, we calculate the rest-frame isotropic radio energy ($E_{\rm rest,400}$) for each FRB.
By integrating Eq. 8 in \citet{Macquart2018b} over the frequency, $E_{\rm rest,400}$ is described as
\begin{equation}
\label{eq:Erest400}
E_{\rm rest,400}=4\pi d_{l}^{2}E_{\rm obs,400}/(1+z),
\end{equation}
where $d_{l}$ is the luminosity distance to the redshift of FRB.
The uncertainty of $E_{\rm rest,400}$ ($\delta E_{\rm rest,400}$) includes the error propagation of $\delta$DM$_{\rm obs}$, $\sigma_{\rm DM_{\rm obs}}$, and the uncertainty of $F_{\nu}$ ($\delta F_{\nu}$) via Eqs. \ref{eq:DM}, \ref{eq:Eobs400_1}, \ref{eq:Eobs400_2}, and \ref{eq:Erest400}.
To estimate $\delta E_{\rm rest,400}$, we performed the same manner as the MC simulations for the redshift uncertainty (see Section \ref{calc_redshift}) with 10,000 iterations.

\subsection{$V_{\rm max}$ and energy function}
\label{vmax}
In this work, we use the $V_{\rm max}$ method \citep[e.g.][]{Schmidt1968,Avni1980} to derive the FRB energy function, i.e., the number density of FRB sources as a function of energy. 
$V_{\rm max}$ is the maximum volume within which each source could still be detected for a certain detection threshold.
The $V_{\rm max}$ method allows us to measure the FRB energy function as it is without any prior assumption on its functional shape.
We follow the method described in \citet{Hashimoto2020a} and \citet{Hashimoto2020c} except for the $z_{\rm max}$ calculation as described in Section \ref{zmax}.
In the following section, we define the sample from which the energy function is derived.

\subsubsection{Sample for the energy function}
\label{sample_EF}
\citet{CHIMEcat2021} reported that significant fractions of FRBs with higher scattering times and lower fluences are missed.
The selection functions for such missing populations are highly uncertain since they are derived from small sample sizes \citep{CHIMEcat2021}.
Therefore, we exclude FRBs with $\log \tau_{\rm scat}>0.8$ (ms) or $\log F_{\nu}<0.5$ (Jy ms) from the calculation of the energy function.
FRBs which satisfy all of the following selection criteria are utilised for the FRB energy function.
\begin{itemize}
\item {\it bonsai\_snr} $>10$
\item DM${\rm obs} >1.5 \times$max(DM$_{\rm NE2001}$, DN$_{\rm YMW16}$)
\item not detected in far side-lobes
\item $\log \tau_{\rm scat} < 0.8$ (ms)
\item {\it excluded\_flag} $=0$
\item the first detected burst if the FRB source is a repeater
\item $\log F_{\nu} >0.5$ (Jy ms)
\end{itemize}
After applying these criteria to the new CHIME catalogue, the selected sample includes in total 176 FRB sources of which 164 are non-repeating FRB sources and 12 are repeating FRB sources.

\subsubsection{Redshift bins}
\label{zbin}
Redshift bins have to be defined for the $V_{\rm max}$ method so that energy functions at different redshifts can be compared.
Because redshift bins are arbitrarily selected, the results may depend on how the redshift bins are selected.
To take this uncertainty into account, two different sets of redshift bins are tested in this work.
One is a set of four redshift bins defined by boundaries at $z=0.05$, 0.30, 0.68, 1,38, and 3.60 for non-repeating FRBs.
These redshifts correspond to the lookback times of 0.7, 3.5, 6.4, 9.2, and 12.1 Gyr, respectively.
Three redshift bins are utilised for repeating FRB sources with boundaries at $z=0.05$, 0.31, 0.72, and 1.50, corresponding to the lookback times of 0.7, 3.6, 6.6, and 9.5 Gyr, respectively.
The interval of redshift bins is decided so that the lookback time between redshift bins is the same for each non-repeating and repeating FRBs.

\begin{figure}
    \includegraphics[width=\columnwidth]{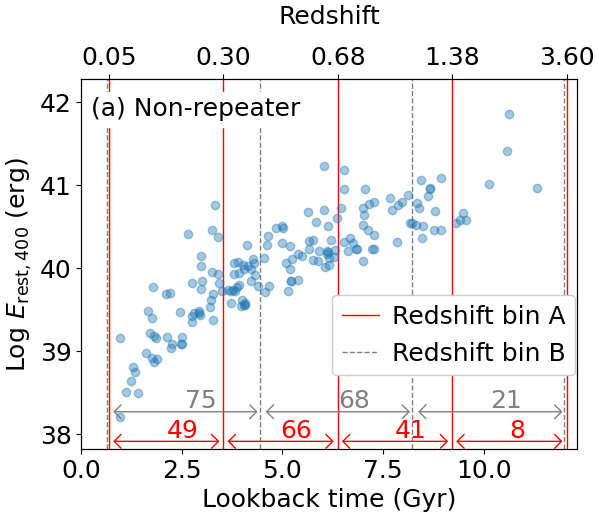}
    \includegraphics[width=\columnwidth]{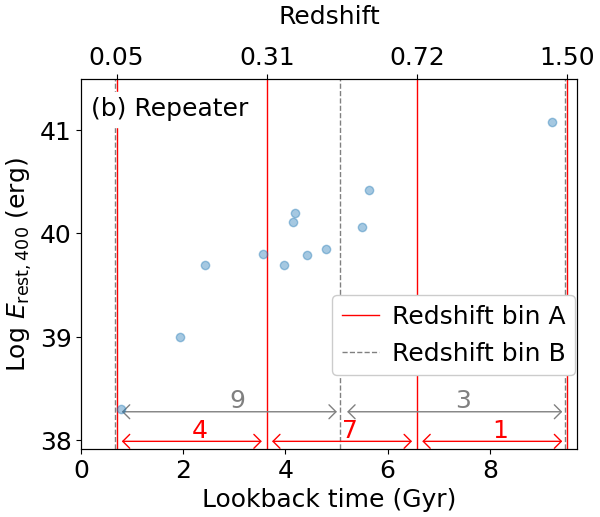}
    \caption{
    The isotropic radio energy integrated over the rest-frame 400 MHz width ($E_{\rm rest,400}$) as a function of lookback time or redshift for (top) non-repeating and (bottom) repeating CHIME FRBs.
    The redshift bins are indicated by vertical lines and horizontal arrows.
    Different colours correspond to the different sets of redshift bins: {\it redshift bin A} shown by red solid lines and {\it redshift bin B} shown by the grey dashed lines.
    The number of samples within each redshift bin is indicated on each arrow.
    The grey dashed lines are slightly shifted for the visualisation purpose.
    }
    \label{fig3}
\end{figure}

Another set of three redshift bins is defined by $z=0.05$, 0.41, 1.09, and 3.60 for non-repeating FRBs and two redshift bins defined by $z=0.05$, 0.49, and 1.50 for repeating FRB sources.
The interval of redshift bins in this case is also defined with the same lookback-time interval for each non-repeating and repeating FRBs.
These redshift bins along with the numbers of sources within bins are summarised in Fig. \ref{fig3}.
We, hereafter, use terminologies of \lq {\it redshift bin A}\rq\ and \lq {\it redshift bin B}\rq\ for red solid lines and grey dashed lines shown in Fig. \ref{fig3}, respectively.

\subsubsection{$z_{\rm max}$ calculation}
\label{zmax}
The 4$\pi$ coverage of $V_{\rm max}$ ($V_{\rm max,4\pi}$) is described as
\begin{equation}
V_{\rm max,4\pi}=\frac{4\pi}{3}(d_{\rm max}^{3}-d_{\rm min}^{3}),
\end{equation}
where $d_{\rm min}$ is the comoving distance to the lower bound of the redshift bin to which an FRB belongs (See Section \ref{zbin} for details) and $d_{\rm max}$ is the maximum comoving distance for the FRB with a certain energy to be detected.
If $d_{\rm max}$ is larger than the comoving distance to the upper bound of the redshift bin, the upper bound distance is utilised as $d_{\rm max}$.
We define $z_{\rm min}$ and $z_{\rm max}$ as corresponding redshifts to $d_{\rm min}$ and ${d_{\rm max}}$, respectively.
Among the selection criteria described in Section \ref{sample_EF}, the SNR and fluence cuts are relevant in our $z_{\rm max}$ calculation since both of them decrease with increasing redshift for each FRB.
We calculate corresponding fluences and SNRs for each FRB at higher redshifts.
At a certain redshift, either the SNR or fluence falls below the criterion, which provides each FRB with $z_{\rm max}$.
Note that the criteria for the dispersion measure and scattering time in Section \ref{sample_EF} are not relevant because the observed dispersion measure and scattering time always satisfy these criteria for higher redshifts.
Here, we assume that the scattering is less significant at higher redshifts due to the $\tau_{\rm scat} \propto \nu_{\rm rest}^{-4}$ dependency.

Using Eq. \ref{eq:Erest400}, the observed fluence of each FRB can be scaled to estimate the corresponding fluence at higher redshifts and $z_{\rm max}$ which satisfies the fluence criterion in Section \ref{sample_EF} ($z_{\rm max,fluence}$).
The notation of \lq fluence\rq\ expresses $z_{\rm max}$ derived from the fluence criterion.
However, the SNR cut is not as simple as the fluence cut because the SNR depends on many factors, e.g., redshift, duration, fluence, detection algorithm, etc.
Therefore, we empirically derive how the SNR scales with redshift.
The SNR should primary depend on $E_{\rm obs}/w_{\rm bc}^{1/2}$ \citep[e.g.][]{Spitler2014,Shannon2018,CHIMEFRB2019}, where $w_{\rm bc}$ is the boxcar duration of FRB \citep[{\it bc\_width} in][]{CHIMEcat2021}.
The boxcar duration represents the observed duration of the FRB pulse including instrumental, scattering, and redshift-broadening effects.

\begin{figure}
    \includegraphics[width=\columnwidth]{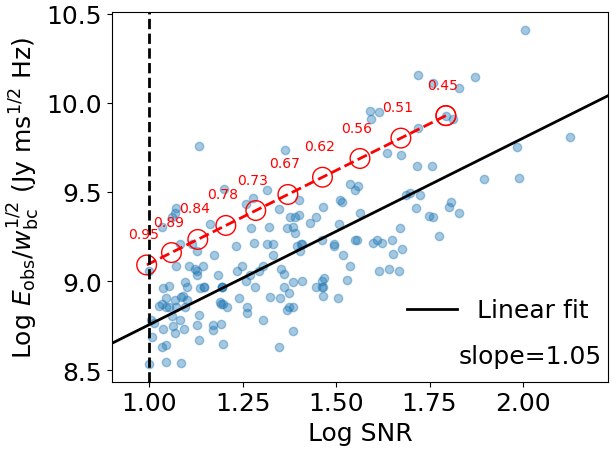}
    \caption{
    $E_{\rm obs}/w_{\rm bc}^{1/2}$ as a function of SNR in the logarithmic scale for the sample described in Section \ref{sample_EF}.
    This sample (blue dots) is utilised for the energy function in Section \ref{results}.
    The black solid line indicates the best-fit linear function to the data.
    The vertical black dashed line is the SNR criterion (SNR$=10$) adopted in this work.
    The red dashed line demonstrates an example of the redshift track of an FRB at $z=0.45$ fixing its isotropic radio energy.
    The open red circles indicate expected $E_{\rm obs}/w_{\rm bc}^{1/2}$ and SNR values of this FRB when different redshifts (labelled numbers) are assumed.
    }
    \label{fig4}
\end{figure}

Fig. \ref{fig4} shows $E_{\rm obs}/w_{\rm bc}^{1/2}$ as a function of SNR in the logarithmic scale for the sample selected in Section \ref{sample_EF}.
In Fig. \ref{fig4}, there is a correlation between $E_{\rm obs}/w_{\rm bc}^{1/2}$ and SNR.
The slope of the best-fit linear function (black solid line) to the data points is $1.05$.
For each FRB, we assume a scaling relation between $E_{\rm obs}/w_{\rm bc}^{1/2}$ and SNR with a linear slope of 1.05 to derive $z_{\rm max, SNR}$.
Here, the notation \lq SNR\rq\ expresses $z_{\rm max}$ derived from the SNR criterion.
An example is demonstrated by red circles on the red dashed line in Fig. \ref{fig4}.
This example selects a particular FRB at $z=0.45$ (the rightmost number labelled for the red circles).
The red circles correspond to $\log(E_{\rm obs}/w_{\rm bc}^{1/2})$ and $\log$(SNR) at different redshifts for this FRB assuming such a scaling relation.
As the redshift increases (numbers labelled for the red circles), corresponding $E_{\rm obs}/w_{\rm bc}^{1/2}$ and SNR decrease.
At $z=0.95$, the corresponding SNR becomes $\log$ SNR $=1.0$, which is the SNR selection criterion. 
This redshift, $z=0.95$, is $z_{\rm max, SNR}$ of this FRB derived by scaling SNR.
In this calculation, Eq. \ref{eq:Erest400} is utilised to scale $E_{\rm obs}$ as a function of redshift.
The redshift dependency of $w_{\rm bc}$ is approximated to be
\begin{equation}
\label{wbcz}
w_{\rm bc,{\it z}} =\sqrt{w_{\rm int,{\it z}}^{2} + (t_{\rm sample}/2.355)^{2} + t_{\rm smear,{\it z}}^{2} + \tau_{\rm scat,{\it z}}^{2}}~({\rm ms}), 
\end{equation}
where 
\begin{equation}
w_{\rm int,{\it z}}=w_{\rm int} \left(\frac{1+z}{1+z_{\rm FRB}}\right)~({\rm ms}),
\end{equation}
\begin{equation}
t_{\rm sample}=0.983~({\rm ms}),
\end{equation}
\begin{equation}
t_{\rm smear,{\it z}}=8.3\times10^{-3} \left( \frac{\rm DM_{\rm obs}}{\rm pc~cm^{-1}} \right) \left( \frac{\Delta\nu_{\rm obs}}{\rm MHz}\right) \left( \frac{\nu_{\rm obs}}{\rm GHz} \right)^{-3}~({\rm ms})
\end{equation}
\begin{equation}
\label{tauz}
\tau_{\rm scat,{\it z}}=\tau_{\rm scat} \left(\frac{1+z_{\rm FRB}}{1+z}\right)^{3}~({\rm ms}).
\end{equation}
We adopt $\Delta\nu_{\rm obs}=24.4\times10^{-3}$ MHz and $\nu_{\rm obs}=0.6$ GHz.
The cubic power in Eq \ref{tauz} is due to the $\tau_{\rm scat}\propto \nu_{\rm rest}^{-4}$ dependency and the $(1+z)$ dependency of the observed time.
When the only upper limit is measured for $w_{\rm int}$ or $\tau_{\rm scat}$, we ignore its term in Eq. \ref{wbcz}.

Both $z_{\rm max, fluence}$ and $z_{\rm max, SNR}$ are calculated for each FRB.
We adopt the smaller one as $z_{\rm max}$.
If the adopted $z_{\rm max}$ is higher than the upper bound of the redshift bin which each FRB belongs to ($z_{\rm upper}$), $z_{\rm upper}$ is used as $z_{\rm max}$, i.e.,
\begin{equation}
z_{\rm max}={\rm min}(z_{\rm max, fluence},z_{\rm max, SNR},z_{\rm upper}).
\end{equation}

\subsubsection{Number density of each FRB source}
\label{calc_density}
\begin{figure*}
    \includegraphics[width=1.8\columnwidth]{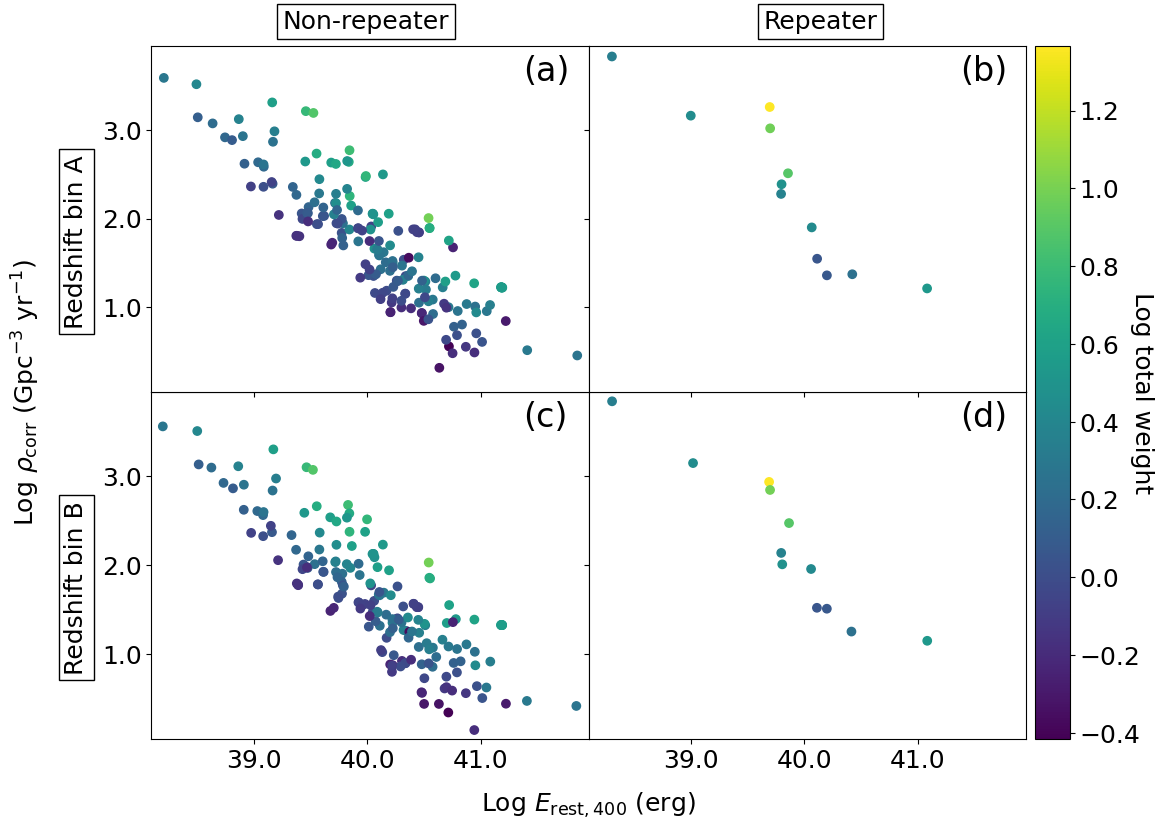}
    \caption{
    The number density of individual CHIME FRBs ($\rho_{\rm corr}$) as a function of isotropic radio energy integrated over the rest-frame 400 MHz width ($E_{\rm rest,400}$).
    The left and right panels show non-repeating and repeating FRBs, respectively.
    The top and bottom panels correspond to {\it redshift bin A} and {\it redshift bin B}, respectively (see Fig. \ref{fig3} for details).
    The number density is corrected for the selection functions described in Section \ref{derived_SF}.
    Colours correspond to the total weight in the logarithmic scale, i.e., $\log [W_{\rm scale}w_{i}({\rm DM})w_{i}(\tau_{\rm scat})w_{i}(w_{\rm int})w_{i}(F_{\nu})]$ for the $i$th FRB.
    }
    \label{fig5}
\end{figure*}

Each FRB was detected in the comoving volume, V$_{\rm max,4\pi} \times \Omega_{\rm sky}$, where $\Omega_{\rm sky}$ is the fractional coverage of the CHIME field of view on the sky.
The number density of each FRB source per unit time, $\rho_{\rm uncorr}(E_{\rm rest,400})$, is 
\begin{equation}
\rho_{\rm uncorr}(E_{\rm rest,400})=\frac{1+z_{\rm FRB}}{V_{\rm max,4\pi}\Omega_{\rm sky}t_{\rm obs}},
\end{equation}
where $t_{\rm obs}$ is the survey time.
We adopt $\Omega_{\rm sky}=0.003$ and $t_{\rm obs}=214.8/365$ = 0.59 year \citep{CHIMEcat2021}.
The subscript \lq uncorr\rq\ indicates that the number density is uncorrected for the observational selection functions described in Section \ref{derived_SF}.
The number density corrected for the selection functions, $\rho_{\rm corr}(E_{\rm rest,400})$, is described as 
\begin{equation}
\rho_{\rm corr}(E_{\rm rest,400})=\rho_{\rm uncorr}W_{\rm scale}w({\rm DM})w(\tau_{\rm scat})w(w_{\rm int})w(F_{\nu}),
\end{equation}
where $w({\rm DM})$, $w(\tau_{\rm scat})$, $w(w_{\rm int})$, and $w(F_{\nu})$ are weight functions derived by the reciprocals of Eqs. \ref{selectionDM}-\ref{selectionfnu}, respectively.
When an FRB shows multiple sub-bursts, the product $\Pi_{\ell=1} w(w_{\rm int,{\it \ell}})$ is adopted as $w(w_{\rm int})$, where the subscript $\ell$ indicates the $\ell$th sub-burst of each FRB.
$W_{\rm scale}$ is the scaling factor for the weight functions.
CHIME's source finding algorithm detected 39,638 sources out of 84,697 injected FRBs \citep{CHIMEcat2021}.
The scaling factor is determined so that the sum of weights over the selected sample matches this fraction, i.e.,
\begin{equation}
W_{\rm scale}\sum_{i=1}^n w_{i}({\rm DM})w_{i}(\tau_{\rm scat})w_{i}(w_{\rm int})w_{i} (F_{\nu})=\frac{84,697}{39,638}\times176,
\end{equation}
where the subscript $i$ denotes the $i$th FRB and 176 is the total number of sample described in Section \ref{sample_EF}.
We note that this scaling factor does not affect our main focus on the redshift evolution of the energy functions and volumetric rates discussed in Section \ref{discussion} because the relative evolution is compared.
Fig. \ref{fig5} shows $\rho_{\rm corr}$ as a function of $E_{\rm rest,400}$ with different colours for the total weights of $\log [W_{\rm scale}w_{i}({\rm DM})w_{i}(\tau_{\rm scat})w_{i}(w_{\rm int})w_{i}(F_{\nu})]$.
The uncertainty of $\rho_{\rm corr}$ ($\delta\rho_{\rm corr}$) is calculated for each FRB by MC simulations with 10,000 iterations, following the same manner as the MC simulations for the uncertainties of redshift and $E_{\rm rest,400}$ (see Section \ref{physpara} for details).

The derived physical parameters including the redshifts, $E_{\rm rest,400}$, weights, and $\rho_{\rm corr}$ are summarised in APPENDIX A.

\subsubsection{Calculation of energy functions}
\label{calc_EF}
Non-repeating and repeating FRBs at each redshift bin are divided into different energy ($E_{\rm rest, 400}$) bins in the logarithmic scale to calculate their energy functions.
Within each energy bin, $\rho_{\rm corr}$ is summed to calculate the energy function ($\Phi$):
\begin{equation}
\phi(z_{\rm median},E_{\rm rest,400,{\it j}})=\sum_{k}\rho_{\rm corr}(E_{\rm rest,400,{\it j,k}})/\Delta\log E_{\rm rest,400},
\end{equation}
where the subscripts $j$ and $k$ mean the $k$th FRB in the $j$th energy bin at each redshift bin with a median redshift of $z_{\rm median}$.
$\Delta \log E_{\rm rest,400}$ is the energy bin size.
The uncertainty of $\Phi$ at each bin is calculated by a quadrature sum of the Poisson uncertainty \citep{Gehrels1986} and $\delta\rho_{\rm corr}$.

\subsubsection{Lower limits of energy functions}
\label{lower_EF}
We excluded FRBs with highly uncertain selection functions in Section \ref{sample_EF} because correcting such selection functions adds huge uncertainties to the derived energy functions.
Instead, we use the excluded sample to estimate lower limits of the energy functions without correcting for the selection functions.
Each FRB in the excluded sample indicates that there is at least one FRB source within the corresponding $V_{\rm max}$, though the actual number of FRBs with the same parameter spaces may be larger due to the selection functions.
This provides each FRB source with a lower limit of the number density.
We performed the analyses described in Section \ref{vmax} for the excluded sample to estimate the lower limits of the energy functions.

\section{Results}
\label{results}
\subsection{Derived energy functions}
\label{derived_EF}
\begin{figure*}
    \includegraphics[width=1.8\columnwidth]{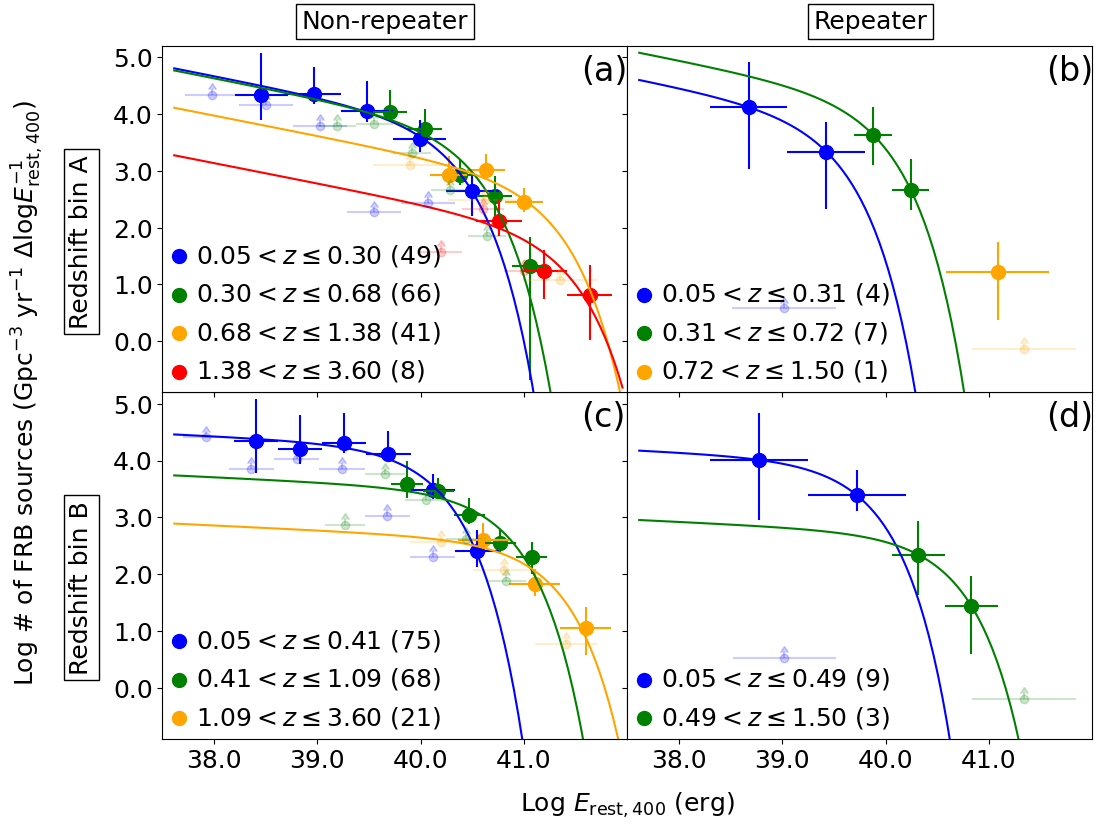}
    \caption{
    The energy functions ($\phi$) of non-repeating and repeating CHIME FRB sources.
    The left and right columns show non-repeating and repeating FRBs, respectively.
    The top and bottom rows correspond to {\it redshift bin A} and {\it redshift bin B}, respectively (see Fig. \ref{fig3} for details).
    The energy functions are corrected for the selection functions described in Section \ref{derived_SF}.
    Colours correspond to the different redshifts.
    The translucent dots with arrows indicate lower limits of the energy functions (see Section \ref{lower_EF} for details).
    The best-fit Schechter functions are shown by solid lines.
    The faint-end slopes of the Schechter functions ($\alpha=-1.4$ for {\it redshift bin A} and $-1.1$ for {\it redshift bin B}) are derived by fitting to the lowest-$z$ bins of non-repeating FRBs.
    These values are respectively assumed for {\it redshift bin A} and {\it redshift bin B} cases of the energy functions of non-repeating FRBs at higher redshift bins and repeating FRB sources.
    }
    \label{fig6}
\end{figure*}

Fig. \ref{fig6} shows the derived energy functions of non-repeating and repeating FRB sources.
At lower redshift bins, the energy functions of non-repeating FRBs show steeper slopes at higher energies while the slopes become flattened towards lower energies, indicating Schechter function-like shapes.
Therefore, we fit Schechter functions to the derived energy functions:
\begin{equation}
\phi(\log E){\rm d}\log E=\phi^{*}\left( \frac{E}{E^{*}} \right)^{\alpha+1}\exp \left(-\frac{E}{E^{*}}\right){\rm d}\log E,
\end{equation}
where $\phi^{*}$ is the normalisation factor, $\alpha+1$ is the faint-end slope, and $E^{*}$ is the break energy of the Schechter function.
Note that $\alpha+1$ is the slope in the logarithmic scale of $\log E_{\rm rest,400}$ whereas $\alpha$ indicates the slope in the linear scale of $E_{\rm rest, 400}$.
We use $E_{\rm rest,400}$ as $E$.
The best-fit faint-end slope is $\alpha=-1.4^{+0.7}_{-0.5}~(-1.1^{+0.6}_{-0.4})$ for non-repeating FRBs at the lowest-$z$ bin of {\it redshift bin A} ({\it redshift bin B}).
Except for the lowest-$z$ bin of non-repeating FRBs, $\alpha$ is poorly constrained due to the lack of data points at lower energies.
Therefore, $\alpha=-1.4~(-1.1)$ is assumed for non-repeating FRBs at higher redshift bins of {\it redshift bin A} ({\it redshift bin B}). 
This is also the case for repeating FRB sources, although their highest-$z$ bin in {\it redshift bin A} case is not fitted with the Schechter function due to lack of data points.
The best-fit parameters of the Schechter functions are summarised in Table \ref{tab1}.
In Fig. \ref{fig6}, the energy functions of non-repeating FRBs show clear decreasing trends (i.e. decrease in $\log$ \# of FRB sources) towards higher redshifts for both redshift bin cases.

The sample size of repeating FRB sources and the number of data points in their energy functions are too small to derive accurate energy functions (Fig. \ref{fig6}b and d). 
As shown in Fig. \ref{fig6} (b) and (d), the redshift evolution of the energy functions of repeating FRB sources strongly depend on the adopted redshift bins and assumed faint end slopes.
More sample of repeating FRB sources is necessary to conclude their energy functions.
Therefore, we leave the further analysis and a discussion about repeating FRB sources for future works.
We focus on non-repeating FRB sources in the following sections unless otherwise mentioned.

\begin{table*}
	\centering
	\caption{
	The best-fit parameters of the Schechter functions to the FRB energy functions.}
	\label{tab1}
	\begin{flushleft}
	\begin{tabular}{|l|c|c|c|c|}\hline
\multicolumn{5}{|c|}{Non-repeating FRBs ({\it redshift bin A})} \\\hline
Redshift bin ($z_{\rm median}$) & $\phi^{*}$ & $E_{\rm rest,400}^{*}$ & $\alpha$ (linear-scale slope) & $\alpha+1$ (logarithmic-scale slope)\\
$0.05<z\leq0.30$ (0.20) & 3.9$^{+0.7}_{-0.6}$ & $40.1^{+0.3}_{-0.3}$ & $-1.4^{+0.7}_{-0.5}$ & $-0.4^{+0.7}_{-0.5}$ \\
$0.30<z\leq0.68$ (0.47) & 3.8$^{+0.5}_{-0.3}$ & $40.3^{+0.2}_{-0.2}$ & $-1.4^{a}$ & $-0.4^{a}$\\
$0.68<z\leq1.38$ (0.94) & 2.9$^{+0.6}_{-6.8}$ & $41.0^{+10.3}_{-0.4}$ & $-1.4^{a}$ & $-0.4^{a}$\\
$0.38<z\leq3.60$ (1.65) & 2.0$^{+0.4}_{-0.8}$ & $41.2^{+0.7}_{-0.3}$ & $-1.4^{a}$ & $-0.4^{a}$\\\hline
\multicolumn{5}{|c|}{Non-repeating FRBs ({\it redshift bin B})} \\\hline
$0.05<z\leq0.41$ (0.25) & 4.3$^{+0.5}_{-0.3}$ & $39.9^{+0.2}_{-0.2}$ & $-1.1^{+0.6}_{-0.4}$ & $-0.1^{+0.6}_{-0.4}$\\
$0.41<z\leq1.09$ (0.64) & 3.5$^{+0.3}_{-0.4}$ & $40.6^{+0.3}_{-0.2}$ & $-1.1^{a}$ & $-0.1^{a}$\\
$1.09<z\leq3.60$ (1.23) & 2.6$^{+0.3}_{-0.4}$ & $41.0^{+0.3}_{-0.2}$ & $-1.1^{a}$ & $-0.1^{a}$\\\hline
\multicolumn{5}{|c|}{ML-selected non-repeating FRBs ({\it redshift bin B}): see APPENDIX B for details} \\\hline
$0.05<z\leq0.41$ (0.24) & 4.0$^{+0.7}_{-7.4}$ & $39.7^{+8.6}_{-0.5}$ & $-1.1^{+1.1}_{-0.8}$ & $-0.1^{+1.1}_{-0.8}$\\
$0.41<z\leq1.09$ (0.64) & 3.4$^{+0.2}_{-0.8}$ & $40.3^{+0.6}_{-0.2}$ & $-1.1^{a}$ & $-0.1^{a}$\\
$1.09<z\leq3.60$ (1.23) & 3.2$^{+0.4}_{-0.8}$ & $40.4^{+0.5}_{-0.3}$ & $-1.1^{a}$ & $-0.1^{a}$\\\hline
\multicolumn{4}{|c|}{Repeating FRB sources ({\it redshift bin A})} \\\hline
$0.05<z\leq0.31$ (0.17) & 4.0$^{+1.2}_{-6.2}$ & $39.3^{+15.3}_{-0.5}$ & $-1.4^{a}$ & $-0.4^{a}$\\
$0.31<z\leq0.72$ (0.4) & 4.3$^{+0.9}_{-1.6}$ & $39.7^{+1.1}_{-0.2}$ & $-1.4^{a}$ & $-0.4^{a}$\\\hline
\multicolumn{4}{|c|}{Repeating FRB sources ({\it redshift bin B})} \\\hline
$0.05<z\leq0.49$ (0.35) & 4.0$^{+1.0}_{-2.1}$ & $39.6^{+13.8}_{-0.4}$ & $-1.1^{a}$ & $-0.1^{a}$\\
$0.49<z\leq1.50$ (0.56) & 2.7$^{+1.0}_{-1.4}$ & $40.4^{+2.7}_{-0.3}$ & $-1.1^{a}$ & $-0.1^{a}$\\\hline
    \end{tabular}\\
    $^{a}$ fixed value.
    \end{flushleft}
\end{table*}

\subsection{Derived volumetric non-repeating FRB rates as a function of redshift}
\label{derived_FRB_rate}
\begin{figure}
    \includegraphics[width=\columnwidth]{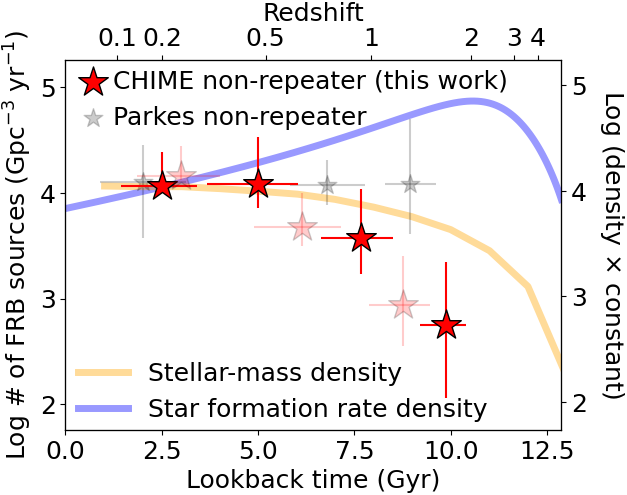}
    \caption{
    The volumetric rate of CHIME non-repeating FRBs as a function of lookback time or redshift.
    The dark red stars and translucent red stars correspond to the volumetric rates derived from {\it redshift bin A} and {\it redshift bin B}, respectively.
    The volumetric rates are calculated by integration of the best-fit Schechter functions to the energy functions.
    The integration range is from $\log E_{\rm rest,400}=39.0$ to 41.5 erg.
    The horizontal coordinate and horizontal error of each data represent the median redshift and the median redshift error of the sample within each redshift bin, respectively.
    The vertical errors are estimated by 10,000 iterations of the MC simulation which take fitting uncertainties to the energy functions (Fig. \ref{fig6}) into account.
    The volumetric rates of Parkes non-repeating FRBs are shown by grey stars \citep{Hashimoto2020c}.
    The cosmic stellar-mass density evolution \citep{Lopez2018} and the cosmic star-formation rate density evolution \citep{Madau2017} are shown by blue and yellow lines, respectively.
    For comparison, these cosmic densities are adjusted at $z=0.20$ such that the cosmic densities and the volumetric rate of CHIME non-repeating FRBs ({\it redshift bin A}) are the same at this redshift.
    The label for the cosmic densities is shown at the right vertical axis, which has the same logarithmic scale as the left axis. 
    }
    \label{fig7}
\end{figure}

To calculate the volumetric non-repeating FRB rates as a function of redshift, the best-fit Schechter function is integrated over $E_{\rm rest,400}=39.0$ to 41.5 erg for each redshift bin.
Fig. \ref{fig7} shows the volumetric rate of non-repeating FRBs as a function of redshift (red stars for {\it redshift bin A} and translucent red stars for {\it redshift bin B}).
The horizontal coordinate and horizontal error of each data (red stars and translucent red stars) represent the median redshift and the median redshift error of the sample within each redshift bin, respectively.
The vertical errors are estimated by 10,000 iterations of the MC simulation which take fitting uncertainties to the energy functions (Fig. \ref{fig6}) into account.
For comparison, the cosmic stellar-mass density evolution \citep[yellow line:][]{Lopez2018} and the cosmic star-formation rate density evolution \citep[blue line:][]{Madau2017} are adjusted at $z=0.20$ such that these cosmic densities and the volumetric rate of CHIME non-repeating FRBs ({\it redshift bin A}) are the same at this redshift.
In Fig. \ref{fig7}, the volumetric non-repeating FRB rates decrease with increasing redshifts for both redshift bin cases.
%This decreasing trend is much closer to the cosmic stellar-mass density evolution than the cosmic star-formation rate density.

To present quantitative differences between the volumetric non-repeating FRB rate and cosmic star-formation rate/stellar-mass densities in Fig. \ref{fig7}, we calculate their $\chi^{2}$ values. 
These density functions are adopted from \citet{Madau2017} and \citet{Lopez2018} \citep[see Eq. 13 and 15 in ][for details]{Hashimoto2020c}, which are shown as blue and orange solid lines in Fig. \ref{fig7}, respectively.
Because the scaling factors between the volumetric non-repeating FRB rate and the cosmic star-formation rate/stellar-mass densities are arbitrary, free constant parameters are added to these density functions in a logarithmic scale. 
We fit these functions to the volumetric non-repeating FRB rates in Fig. \ref{fig7}. 
The sum of squared deviations weighted by uncertainties is adopted as the $\chi^{2}$ value. 

The $\chi^{2}$ values to the cosmic stellar-mass density are 2.4 and 5.3 for {\it redshift bin A} and {\it redshift bin B}, respectively. 
The corresponding $p$-values are 0.11 and 0.02, indicating that the null hypothesis is not rejected with a 1\% significance threshold. 
The $\chi^{2}$ values to the cosmic star-formation rate density are 13.5 and 19.5 for {\it redshift bin A} and {\it redshift bin B}, respectively. 
The corresponding $p$-values are 2.4e$-4$ and 9.8e$-6$, indicating that the null hypothesis is ruled out with a more than 99\% confidence level. 
In summary, there is no significant difference between the volumetric non-repeating FRB rate and cosmic stellar-mass density evolution with the 1\% significance threshold, whereas the difference to the cosmic star-formation rate density is statistically significant.

\section{Discussion}
\label{discussion}
\subsection{Luminosity function and energy function}
FRB luminosity functions and energy functions have been investigated in previous works. 
Since the absolute values of the FRB luminosity and energy functions depend on how the FRB selection functions are scaled (see Section \ref{calc_density} for details), we discuss their slopes in this work.
\citet{Luo2020} reported the FRB luminosity function using a total of 46 FRBs.
They used flux densities to calculate isotropic FRB radio luminosities in units of erg s$^{-1}$.
The reported slope of the luminosity function is $\alpha_{\rm Sc}=-1.79^{+0.31}_{-0.35}$, where the subscript \lq Sc\rq\ denotes the slope of a Schechter function.
Their sample is heterogeneous because (i) the FRBs collected from Parkes, Arecibo, Green Bank Telescope, UTMOST, and ASKAP are all mixed and (ii) non-repeating and repeating FRBs are also combined.
To mitigate systematics between different telescopes, \citet{Hashimoto2020a} and \citet{Hashimoto2020c} unified the definition of the detection threshold for different telescopes and estimated the time-integrated luminosity function for each telescope and each of non-repeating and repeating FRB sources.
They adopted the time-integrated luminosity in units of erg Hz$^{-1}$ using fluences because the fluence is less affected by the observational time resolution compared with the flux density \citep{Macquart2018}.
The best-fitting power-law slope is $\alpha_{\rm PL}=-1.35^{+0.51}_{-0.50}$ in the linear scale for Parkes non-repeating FRBs at $0.01<z\leq 0.35$ \citep{Hashimoto2020c}.
The subscript \lq PL\rq\ expresses the power-law slope.
The slopes in these works \citep{Luo2020,Hashimoto2020a,Hashimoto2020c} are consistent with each other within the errors, although they adopted different approaches to investigate the FRB luminosity functions.

In this work, we update the work by \citet{Hashimoto2020a} using the new CHIME FRBs \citep{CHIMEcat2021}.
The energy functions of non-repeating FRBs are derived from 164 FRBs in this work, which is about one order of magnitude larger than that in \citet{Hashimoto2020c} (23 Parkes non-repeating FRBs).
Our sample is homogeneous because FRBs are detected by the identical survey instrument, CHIME.
The measured slopes of energy functions of non-repeating FRBs are $-1.4^{+0.7}_{-0.5}$ and $-1.1^{+0.6}_{-0.4}$ for {\it redshift bin A} and {\it redshift bin B}, respectively (Table \ref{tab1}).
These values are consistent with those derived by \citet{Luo2020} and \citet{Hashimoto2020c} within errors.

\citet{Hashimoto2020c} empirically utilised power-law functions to fit the derived time-integrated luminosity functions.
This is because the number of Parkes FRBs is not enough to include more parameters in the fitting functions.
The large sample size of the new CHIME data allows us to investigate the shape of the FRB energy function.
The faint-end slope of the energy functions is significantly flatter than that at the bright end regardless of the redshift bins as shown in Fig. \ref{fig6}.
This shape of the energy function can be reasonably fitted with the Schechter functions.
The Schechter functions have been conventionally assumed in previous works \citep[e.g.][]{Luo2018,Luo2020}.
We observationally confirmed the Schechter function-like shape works, placing an anchor for further investigation of FRB populations in the future.

\subsection{Redshift evolution of luminosity/energy functions and volumetric FRB rate}
\label{volumetric_FRBrate}
The redshift evolution of the FRB luminosity/energy function provides an important implication on the FRB origin \citep[e.g.][]{Locatelli2019,Hashimoto2020c,Arcus2021,James2022}.
For instance, the redshift evolution should follow the cosmic star-formation rate density if FRB progenitors are related to star-forming activities.

There are mainly two approaches to the redshift evolution of FRB luminosity/energy functions.
One is to parameterise the FRB population with the redshift evolution term, dispersion measure, spectral shape, and duration to construct FRB population models.  
To find the best-fit parameters, such models can be fitted to the observed distributions of dispersion measure or other observable using the observational detection threshold.
Based on such an approach, \citet{Arcus2021} used ASKAP and Parkes FRBs to show that both cases of (i) evolution proportional to the cosmic star formation rate and (ii) no evolution can explain the observed distribution of dispersion measures depending on the spectral index and the slope of the luminosity function.
\citet{James2022} assumed the FRB population scaling with the cosmic star-formation rate density with a free power-law parameter.
Under this assumption, they concluded that the FRBs evolve in the same manner as, or faster than, the cosmic star-formation rate density using 24 ASKAP and 20 Parkes FRBs. 
However, the prior assumptions on the redshift evolution made in this approach \citep[e.g.][]{Arcus2021,James2022} may affect the resulting conclusions about the redshift evolution.
\citet{Arcus2021} utilised a mixed sample including both non-repeating and repeating FRB sources \citep[e.g. repeating FRB 171019;][]{Kumar2019}, though non-repeating FRBs are the majority in their sample.

Another approach is to empirically derive the FRB number density as a function of redshift.
This approach is free from such prior assumptions on the redshift evolution of the FRB population, which in principle allows us to measure the redshift dependency as it is.
\citet{Locatelli2019} performed the so-called $\left<V/V_{\rm max} \right>$ test \citep{Schmidt1968} using 23 ASKAP and 20 Parkes FRBs, where $V$ is the volume within which each FRB source is distributed and $V_{\rm max}$ is the maximum volume within which the source could be detected under a certain detection threshold.
They presented that the $\left<V/V_{\rm max} \right>$ of Parkes FRB is consistent with that predicted from the evolution proportional to the cosmic star-formation rate density while the ASKAP FRBs suggests faster evolution towards higher redshifts.
However, comparing the absolute values of $\left<V/V_{\rm max} \right>$ with models might not be straightforward because of the complicated detection threshold and selection functions depending on telescopes \citep[e.g.][]{CHIMEcat2021}.

To mitigate the systematic uncertainty of each telescope, \citet{Hashimoto2020c} used the $V_{\rm max}$ method and time-integrated luminosity functions to investigate the relative number evolution of FRBs within the homogeneous sample.
They used 23 Parkes FRBs to conclude that there is no significant redshift evolution of the volumetric non-repeating FRB rate up to $z\sim2$.
Repeating FRB population may similarly increase towards higher redshifts to the cosmic star-formation rate density if the slopes of their time-integrated luminosity functions at high-$z$ are the same as that at low-$z$ \citep{Hashimoto2020c}.
This work is also treating repeating and non-repeating FRBs separately.

In this work, we update the relative redshift evolution of the non-repeating FRB population tested by \citet{Hashimoto2020c} using the new CHIME FRBs.
This is about one order of magnitude larger homogeneous sample than before.
The energy functions (Fig. \ref{fig6}) and the volumetric rate of CHIME non-repeating FRBs (Fig. \ref{fig7}) show clear decreasing trends towards higher redshifts regardless of how the redshift bins are defined.
%The decreasing trend is much closer to the cosmic stellar-mass density evolution (the yellow line in Fig. \ref{fig7}) than the star-formation rate density evolution (the blue line in Fig. \ref{fig7}).
There is no significant difference between the volumetric non-repeating FRB rate and cosmic stellar-mass density evolution with the 1\% significance threshold, whereas the difference to the cosmic star-formation rate density is statistically significant (see Section \ref{derived_FRB_rate} for details).
This result is consistent with that by \citet{Hashimoto2020c} in the sense that the cosmic stellar-mass density evolution is likely controlling the non-repeating FRB rate, suggesting the old populations as the origin of non-repeating FRBs such as (old) neutron stars and black holes.

The decreasing trend of the volumetric non-repeating FRB rate is %inconsistent to 
apparently different from that found by \citet{James2022}.
Two possible reasons are as follows.
One is the improved sample in this work. 
\citet{James2022} mixed two different samples from ASKAP and Parkes that show different redshift distributions \citep[e.g.][]{Locatelli2019,Hashimoto2020c} and fit models to them all together, whereas we use the homogeneous CHIME sample with better statistics of 176 FRBs.
Another reason would be the difference in approaches.
The model fitting approaches \citep[e.g.][]{James2022,Arcus2021} parameterise the redshift evolution term. 
However, the actual redshift evolution might not fit in the framework of parameterisation.
For instance, a significant decreasing trend of the FRB population towards higher redshift might be difficult to reproduce once the evolution is assumed to be scaled with the cosmic star-formation rate density.
\citet{James2022} and \citet{Arcus2021} do not test old population models.
In contrast, our results do not include any prior assumption on the redshift evolution of the FRB population but show the redshift evolution as measured. 

\citet{Zhangetal2021} tested the observed Parkes and ASKAP samples on redshift distribution models tracking the two extremes of evolving redshift distribution models (star-formation rate history and compact binary merger history).
They concluded that the limited data sample was consistent with both of those models.
The compact binary merger history, i.e., old population scenario, is not ruled out by \citet{Zhangetal2021}.
Therefore, our result is consistent with that found by \citet{Zhangetal2021}.

\citet{Zhang2021} extended prior assumptions on the redshift evolution to the cosmic stellar-mass density evolution and the combination between young and old stellar populations.
They found it difficult for the cosmic star-formation rate density scenario to simultaneously reproduce the observed distributions of dispersion measures and fluences of the new CHIME FRBs.
The best-fit scenario is either (i) a significant delay of FRBs ($\sim 10$ Gyr) with respect to star formation or (ii) a hybrid model of young and old stellar populations with the dominant component of the old stellar population.
Because both (i) and (ii) are dominated by old stellar populations, their results support our hypothesis of old populations as the origin of non-repeating FRBs.
We note that \citet{Zhang2021} used a 5\% significance threshold to conclude that the energy and dispersion measure distributions of the new CHIME FRBs are not consistent with those predicted from the pure cosmic stellar-mass density evolution. 
%However, the 10\% threshold is higher than the standard thresholds of 1\% or 5\% used in statistical tests. 
%Using such a high threshold might be misleading. 
In this work, we use a threshold of 1\% to be more conservative. 
%\textcolor{cyan}{[SY: It might be interesting to predict a rough rate of non-repeating FRBs at the very faint end (e.g., $E_{\rm CHIME}\sim3\times10^{34}$ erg for Galactic FRB 200428) in our Galaxy based on your results. Please neglect it if you find this unimportant.]}

\subsection{Minimising the possible contamination of repeating FRB sources}
More or less, non-repeating FRB sources are likely contaminated by repeating FRB sources because repeating FRBs may be missed by the limited observational time and limited sensitivities, which could lead to the misclassification of non-repeating FRBs. 
\citet{Chen2022} utilised an unsupervised machine learning approach to classify the new CHIME FRBs.
They found nine clustered groups of FRBs, among which four groups of non-repeating FRBs \citep[{\it other\_cluster\_1}, {\it other\_cluster\_2}, {\it other\_cluster\_4}, and {\it other\_cluster\_6} in][]{Chen2022} do not include any repeating FRBs.
Such non-repeating FRBs are expected to be less contaminated by repeating FRB sources.
In APPENDIX B, we use the four groups of non-repeating FRBs to derive their energy functions and volumetric rates.
We found that the derived energy functions and volumetric rates also show a similar trend to Figs. \ref{fig6} and \ref{fig7}: the energy functions and volumetric rates of non-repeating FRB decrease towards higher redshifts (see APPENDIX B for details). 
This suggests that the possible contamination of repeating FRB sources in the non-repeating FRB sample does not significantly affect our results presented in Section \ref{results}.

\subsection{Missing FRB population}
\label{missing_pop}
As described in Section \ref{lower_EF}, we excluded FRBs with highly uncertain selection functions.
The excluded FRBs show high scattering time ($\log \tau_{\rm scat} >0.8$ ms) or low fluence ($\log F_{\nu}<0.5$ Jy ms).
The selection functions of these FRBs are very small, and thus the weighting factors to correct for the selection could be very high ($\gtrsim 20$), though their uncertainties are large.
Therefore, such missing FRB populations may have a significant effect on the energy function.
\citet{CHIMEcat2021} found that there could be a significant fraction of FRBs at $\log \tau_{\rm scat}>1.0$ ms where the CHIME detection algorithm becomes insensitive.
The energy functions and volumetric FRB rates presented in this work are based on the well-explored FRB populations with robust measurements and known selection functions.
We note that energy functions and volumetric rates of such missing FRB populations might show different redshift evolution from our results.
Expanding FRB samples to include large scattering times and lower fluences are important to address the missing FRB population and to fully understand the FRB progenitors. 

\section{Conclusion}
\label{conclusion}
The FRB energy function (the number density of FRB sources as a function of isotropic radio energy) and their redshift evolution provide us with a strong clue to constrain possible FRB progenitors statistically.
If the FRB progenitor is related to young stellar populations, and thus, to the star-formation activity, the energy functions and the volumetric FRB rates should increase towards higher redshifts because the cosmic star-formation rate density increases with increasing redshift up to $z\sim2$.
In contrast, the energy functions and the volumetric FRB rates are expected to decrease towards higher redshifts if old populations such as old neutron stars and black holes are predominantly responsible for the FRB mechanism.
Previous works suffer from small sample sizes and the heterogeneity of the FRB samples (i.e., FRB samples collected from different telescopes), which hampers inferring a definite conclusion.
The new CHIME FRBs are homogeneous in the sense that they are detected with the identical instrument and sensitivity.
The new CHIME FRB catalogue allows us to overcome this problem with a much larger and still homogeneous FRB sample.
We use 164 non-repeating FRB sources and 12 repeating FRB sources selected from the new CHIME catalogue to derive their energy functions.
The non-repeating FRBs in this work are about one order of magnitude larger homogeneous sample than those in previous works.

In this work, the $V_{\rm max}$ method is adopted, which allows us to measure the redshift evolution of the energy functions as it is without any prior assumption on the evolution, unlike model-fitting approaches. 
We find that the energy functions of non-repeating FRBs show Schechter function-like shapes to at least $z\lesssim1$.
The volumetric FRB rates are derived from the integration of the energy functions over the energy.
Both the energy functions and volumetric rate of non-repeating FRBs show a clear decreasing trend towards higher redshifts up to $z\sim2$.
This decreasing trend is more similar to that of the cosmic stellar-mass density evolution than the cosmic star-formation rate density evolution, suggesting old populations as the origin of the majority of non-repeating FRBs:
there is no significant difference between the volumetric non-repeating FRB rate and cosmic stellar-mass density evolution with the 1\% significance threshold, whereas the cosmic star-formation rate scenario is rejected with a more than 99\% confidence level.
These results are based on the well-explored FRBs with robust measurements and known selection functions. 
Our sample, therefore, does not include possible missing FRB populations with large scattering time or lower fluence due to the large uncertainties of selection functions.
Such missing FRB populations have to be investigated via future observations to fully understand the redshift evolution of the FRB energy functions and volumetric rates.

\section*{Acknowledgements}
We are very grateful to the anonymous referee for many insightful comments.
%We thank Dr. Susumu Inoue and Dr. Shintaro Yoshiura for useful discussions.
%AYLO is supported by the Centre for Informatics and Computation in Astronomy (CICA) at National Tsing Hua University (NTHU) through a grant from the Ministry of Education of Taiwan.
TG and TH acknowledge the supports of the Ministry of Science and Technology of Taiwan through grants 108-2628-M-007-004-MY3 and 110-2112-M-005-013-MY3, respectively. 
SY is supported by the advanced ERC grant TReX.
%AYLO's visit to NTHU was supported by the Ministry of Science and Technology of the ROC (Taiwan) grant 105-2119-M-007-028-MY3, hosted by Prof. Albert Kong.
This work used high-performance computing facilities operated by the Centre for Informatics and Computation in Astronomy (CICA) at National Tsing Hua University (NTHU). 
This equipment was funded by the Ministry of Education of Taiwan, the Ministry of Science and Technology of Taiwan, and NTHU.
This research has used NASA's Astrophysics Data System.

%%%%%%%%%%%%%%%%%%%%%%%%%%%%%%%%%%%%%%%%%%%%%%%%%%
\section*{Data availability}
The original CHIME FRB data underlying this article is available at \url{https://www.chime-frb.ca/catalog}.
The parameters derived in this work are available at the article online supplementary material.


\bibliographystyle{mnras}
%\bibliography{CHIME_LF_mnras} % if your bibtex file is called example.bib

\newcommand{\noopsort}[1]{}
\begin{thebibliography}{}
\makeatletter
\relax
\def\mn@urlcharsother{\let\do\@makeother \do\$\do\&\do\#\do\^\do\_\do\%\do\~}
\def\mn@doi{\begingroup\mn@urlcharsother \@ifnextchar [ {\mn@doi@}
  {\mn@doi@[]}}
\def\mn@doi@[#1]#2{\def\@tempa{#1}\ifx\@tempa\@empty \href
  {http://dx.doi.org/#2} {doi:#2}\else \href {http://dx.doi.org/#2} {#1}\fi
  \endgroup}
\def\mn@eprint#1#2{\mn@eprint@#1:#2::\@nil}
\def\mn@eprint@arXiv#1{\href {http://arxiv.org/abs/#1} {{\tt arXiv:#1}}}
\def\mn@eprint@dblp#1{\href {http://dblp.uni-trier.de/rec/bibtex/#1.xml}
  {dblp:#1}}
\def\mn@eprint@#1:#2:#3:#4\@nil{\def\@tempa {#1}\def\@tempb {#2}\def\@tempc
  {#3}\ifx \@tempc \@empty \let \@tempc \@tempb \let \@tempb \@tempa \fi \ifx
  \@tempb \@empty \def\@tempb {arXiv}\fi \@ifundefined
  {mn@eprint@\@tempb}{\@tempb:\@tempc}{\expandafter \expandafter \csname
  mn@eprint@\@tempb\endcsname \expandafter{\@tempc}}}

\bibitem[\protect\citeauthoryear{{Arcus}, {Macquart}, {Sammons}, {James}  \&
  {Ekers}}{{Arcus} et~al.}{2021}]{Arcus2021}
{Arcus} W.~R.,  {Macquart} J.~P.,  {Sammons} M.~W.,  {James} C.~W.,   {Ekers}
  R.~D.,  2021, \mn@doi [\mnras] {10.1093/mnras/staa3948}, \href
  {https://ui.adsabs.harvard.edu/abs/2021MNRAS.501.5319A} {501, 5319}

\bibitem[\protect\citeauthoryear{{Avni} \& {Bahcall}}{{Avni} \&
  {Bahcall}}{1980}]{Avni1980}
{Avni} Y.,  {Bahcall} J.~N.,  1980, \mn@doi [\apj] {10.1086/157673}, \href
  {https://ui.adsabs.harvard.edu/abs/1980ApJ...235..694A} {235, 694}

\bibitem[\protect\citeauthoryear{{Bhandari} et~al.,}{{Bhandari}
  et~al.}{2020}]{Bhandari2020}
{Bhandari} S.,  et~al., 2020, \mn@doi [\apjl] {10.3847/2041-8213/ab672e}, \href
  {https://ui.adsabs.harvard.edu/abs/2020ApJ...895L..37B} {895, L37}

\bibitem[\protect\citeauthoryear{{Bhandari} et~al.,}{{Bhandari}
  et~al.}{2021}]{Bhandari2021}
{Bhandari} S.,  et~al., 2021, arXiv e-prints, \href
  {https://ui.adsabs.harvard.edu/abs/2021arXiv210801282B} {p. arXiv:2108.01282}

\bibitem[\protect\citeauthoryear{{Bhardwaj} et~al.,}{{Bhardwaj}
  et~al.}{2021}]{Bhardwaj2021}
{Bhardwaj} M.,  et~al., 2021, \mn@doi [\apjl] {10.3847/2041-8213/abeaa6}, \href
  {https://ui.adsabs.harvard.edu/abs/2021ApJ...910L..18B} {910, L18}

\bibitem[\protect\citeauthoryear{{Bochenek}, {Kulkarni}, {Ravi}, {McKenna},
  {Hallinan}  \& {Belov}}{{Bochenek} et~al.}{2020}]{Bochenek2020}
{Bochenek} C.,  {Kulkarni} S.,  {Ravi} V.,  {McKenna} D.,  {Hallinan} G.,
  {Belov} K.,  2020, The Astronomer's Telegram, \href
  {https://ui.adsabs.harvard.edu/abs/2020ATel13684....1B} {13684, 1}

\bibitem[\protect\citeauthoryear{{CHIME/FRB Collaboration} et~al.,}{{CHIME/FRB
  Collaboration} et~al.}{2019}]{CHIMEFRB2019}
{CHIME/FRB Collaboration} et~al., 2019, \mn@doi [\nat]
  {10.1038/s41586-018-0867-7}, \href
  {https://ui.adsabs.harvard.edu/abs/2019Natur.566..230C} {566, 230}

\bibitem[\protect\citeauthoryear{{Chen}, {Hashimoto}, {Goto}, {Kim}, {Santos},
  {On}, {Lu}  \& {Hsiao}}{{Chen} et~al.}{2022}]{Chen2022}
{Chen} B.~H.,  {Hashimoto} T.,  {Goto} T.,  {Kim} S.~J.,  {Santos} D. J.~D.,
  {On} A. Y.~L.,  {Lu} T.-Y.,   {Hsiao} T. Y.~Y.,  2022, \mn@doi [\mnras]
  {10.1093/mnras/stab2994}, \href
  {https://ui.adsabs.harvard.edu/abs/2022MNRAS.509.1227C} {509, 1227}

\bibitem[\protect\citeauthoryear{{Dolag}, {Gaensler}, {Beck}  \&
  {Beck}}{{Dolag} et~al.}{2015}]{Dolag2015}
{Dolag} K.,  {Gaensler} B.~M.,  {Beck} A.~M.,   {Beck} M.~C.,  2015, \mn@doi
  [\mnras] {10.1093/mnras/stv1190}, \href
  {https://ui.adsabs.harvard.edu/abs/2015MNRAS.451.4277D} {451, 4277}

\bibitem[\protect\citeauthoryear{{Gehrels}}{{Gehrels}}{1986}]{Gehrels1986}
{Gehrels} N.,  1986, \mn@doi [\apj] {10.1086/164079}, \href
  {https://ui.adsabs.harvard.edu/abs/1986ApJ...303..336G} {303, 336}

\bibitem[\protect\citeauthoryear{{Hashimoto}, {Goto}, {Wang}, {Kim}, {Ho},
  {On}, {Lu}  \& {Santos}}{{Hashimoto} et~al.}{2020a}]{Hashimoto2020a}
{Hashimoto} T.,  {Goto} T.,  {Wang} T.-W.,  {Kim} S.~J.,  {Ho} S. C.~C.,  {On}
  A. Y.~L.,  {Lu} T.-Y.,   {Santos} D. J.~D.,  2020a, \mn@doi [\mnras]
  {10.1093/mnras/staa895}, \href
  {https://ui.adsabs.harvard.edu/abs/2020MNRAS.494.2886H} {494, 2886}

\bibitem[\protect\citeauthoryear{{Hashimoto} et~al.,}{{Hashimoto}
  et~al.}{2020b}]{Hashimoto2020c}
{Hashimoto} T.,  et~al., 2020b, \mn@doi [\mnras] {10.1093/mnras/staa2490},
  \href {https://ui.adsabs.harvard.edu/abs/2020MNRAS.498.3927H} {498, 3927}

\bibitem[\protect\citeauthoryear{{James}, {Prochaska}, {Macquart},
  {North-Hickey}, {Bannister}  \& {Dunning}}{{James} et~al.}{2022}]{James2022}
{James} C.~W.,  {Prochaska} J.~X.,  {Macquart} J.~P.,  {North-Hickey} F.~O.,
  {Bannister} K.~W.,   {Dunning} A.,  2022, \mn@doi [\mnras]
  {10.1093/mnrasl/slab117}, \href
  {https://ui.adsabs.harvard.edu/abs/2022MNRAS.510L..18J} {510, L18}

\bibitem[\protect\citeauthoryear{{Kirsten} et~al.,}{{Kirsten}
  et~al.}{2021a}]{Kirsten2021}
{Kirsten} F.,  et~al., 2021a, arXiv e-prints, \href
  {https://ui.adsabs.harvard.edu/abs/2021arXiv210511445K} {p. arXiv:2105.11445}

\bibitem[\protect\citeauthoryear{{Kirsten}, {Snelders}, {Jenkins}, {Nimmo},
  {van den Eijnden}, {Hessels}, {Gawro{\'n}ski}  \& {Yang}}{{Kirsten}
  et~al.}{2021b}]{Kirsten2020}
{Kirsten} F.,  {Snelders} M.~P.,  {Jenkins} M.,  {Nimmo} K.,  {van den Eijnden}
  J.,  {Hessels} J.~W.~T.,  {Gawro{\'n}ski} M.~P.,   {Yang} J.,  2021b, \mn@doi
  [Nature Astronomy] {10.1038/s41550-020-01246-3}, \href
  {https://ui.adsabs.harvard.edu/abs/2021NatAs...5..414K} {5, 414}

\bibitem[\protect\citeauthoryear{{Kumar} et~al.,}{{Kumar}
  et~al.}{2019}]{Kumar2019}
{Kumar} P.,  et~al., 2019, \mn@doi [\apjl] {10.3847/2041-8213/ab5b08}, \href
  {https://ui.adsabs.harvard.edu/abs/2019ApJ...887L..30K} {887, L30}

\bibitem[\protect\citeauthoryear{{Li}, {Gao}, {Wei}, {Yang}, {Zhang}  \&
  {Zhu}}{{Li} et~al.}{2020}]{Li2020}
{Li} Z.,  {Gao} H.,  {Wei} J.~J.,  {Yang} Y.~P.,  {Zhang} B.,   {Zhu} Z.~H.,
  2020, \mn@doi [\mnras] {10.1093/mnrasl/slaa070}, \href
  {https://ui.adsabs.harvard.edu/abs/2020MNRAS.496L..28L} {496, L28}

\bibitem[\protect\citeauthoryear{{Locatelli}, {Ronchi}, {Ghirlanda}  \&
  {Ghisellini}}{{Locatelli} et~al.}{2019}]{Locatelli2019}
{Locatelli} N.,  {Ronchi} M.,  {Ghirlanda} G.,   {Ghisellini} G.,  2019,
  \mn@doi [\aap] {10.1051/0004-6361/201834722}, \href
  {https://ui.adsabs.harvard.edu/abs/2019A&A...625A.109L} {625, A109}

\bibitem[\protect\citeauthoryear{{L{\'o}pez Fern{\'a}ndez} et~al.,}{{L{\'o}pez
  Fern{\'a}ndez} et~al.}{2018}]{Lopez2018}
{L{\'o}pez Fern{\'a}ndez} R.,  et~al., 2018, \mn@doi [\aap]
  {10.1051/0004-6361/201732358}, \href
  {https://ui.adsabs.harvard.edu/abs/2018A&A...615A..27L} {615, A27}

\bibitem[\protect\citeauthoryear{{Lorimer}}{{Lorimer}}{2021}]{Lorimer2021}
{Lorimer} D.,  2021, \mn@doi [Nature Astronomy] {10.1038/s41550-021-01465-2},
  \href {https://ui.adsabs.harvard.edu/abs/2021NatAs...5..870L} {5, 870}

\bibitem[\protect\citeauthoryear{{Lorimer}, {Bailes}, {McLaughlin}, {Narkevic}
  \& {Crawford}}{{Lorimer} et~al.}{2007}]{Lorimer2007}
{Lorimer} D.~R.,  {Bailes} M.,  {McLaughlin} M.~A.,  {Narkevic} D.~J.,
  {Crawford} F.,  2007, \mn@doi [Science] {10.1126/science.1147532}, \href
  {https://ui.adsabs.harvard.edu/abs/2007Sci...318..777L} {318, 777}

\bibitem[\protect\citeauthoryear{{Luo}, {Lee}, {Lorimer}  \& {Zhang}}{{Luo}
  et~al.}{2018}]{Luo2018}
{Luo} R.,  {Lee} K.,  {Lorimer} D.~R.,   {Zhang} B.,  2018, \mn@doi [\mnras]
  {10.1093/mnras/sty2364}, \href
  {https://ui.adsabs.harvard.edu/abs/2018MNRAS.481.2320L} {481, 2320}

\bibitem[\protect\citeauthoryear{{Luo}, {Men}, {Lee}, {Wang}, {Lorimer}  \&
  {Zhang}}{{Luo} et~al.}{2020}]{Luo2020}
{Luo} R.,  {Men} Y.,  {Lee} K.,  {Wang} W.,  {Lorimer} D.~R.,   {Zhang} B.,
  2020, \mn@doi [\mnras] {10.1093/mnras/staa704}, \href
  {https://ui.adsabs.harvard.edu/abs/2020MNRAS.tmp..667L} {}

\bibitem[\protect\citeauthoryear{{Macquart} \& {Ekers}}{{Macquart} \&
  {Ekers}}{2018a}]{Macquart2018}
{Macquart} J.~P.,  {Ekers} R.~D.,  2018a, \mn@doi [\mnras]
  {10.1093/mnras/stx2825}, \href
  {https://ui.adsabs.harvard.edu/abs/2018MNRAS.474.1900M} {474, 1900}

\bibitem[\protect\citeauthoryear{{Macquart} \& {Ekers}}{{Macquart} \&
  {Ekers}}{2018b}]{Macquart2018b}
{Macquart} J.~P.,  {Ekers} R.,  2018b, \mn@doi [\mnras]
  {10.1093/mnras/sty2083}, \href
  {https://ui.adsabs.harvard.edu/abs/2018MNRAS.480.4211M} {480, 4211}

\bibitem[\protect\citeauthoryear{{Macquart} et~al.,}{{Macquart}
  et~al.}{2020}]{Macquart2020}
{Macquart} J.~P.,  et~al., 2020, \mn@doi [\nat] {10.1038/s41586-020-2300-2},
  \href {https://ui.adsabs.harvard.edu/abs/2020Natur.581..391M} {581, 391}

\bibitem[\protect\citeauthoryear{{Madau} \& {Dickinson}}{{Madau} \&
  {Dickinson}}{2014}]{Madau2014}
{Madau} P.,  {Dickinson} M.,  2014, \mn@doi [\araa]
  {10.1146/annurev-astro-081811-125615}, \href
  {https://ui.adsabs.harvard.edu/abs/2014ARA&A..52..415M} {52, 415}

\bibitem[\protect\citeauthoryear{{Madau} \& {Fragos}}{{Madau} \&
  {Fragos}}{2017}]{Madau2017}
{Madau} P.,  {Fragos} T.,  2017, \mn@doi [\apj] {10.3847/1538-4357/aa6af9},
  \href {https://ui.adsabs.harvard.edu/abs/2017ApJ...840...39M} {840, 39}

\bibitem[\protect\citeauthoryear{{Marcote} et~al.,}{{Marcote}
  et~al.}{2020}]{Marcote2020}
{Marcote} B.,  et~al., 2020, \mn@doi [\nat] {10.1038/s41586-019-1866-z}, \href
  {https://ui.adsabs.harvard.edu/abs/2020Natur.577..190M} {577, 190}

\bibitem[\protect\citeauthoryear{{Petroff} et~al.,}{{Petroff}
  et~al.}{2016}]{Petroff2016}
{Petroff} E.,  et~al., 2016, \mn@doi [\pasa] {10.1017/pasa.2016.35}, \href
  {https://ui.adsabs.harvard.edu/abs/2016PASA...33...45P} {33, e045}

\bibitem[\protect\citeauthoryear{{Piro} et~al.,}{{Piro}
  et~al.}{2021}]{Piro2021}
{Piro} L.,  et~al., 2021, \mn@doi [\aap] {10.1051/0004-6361/202141903}, \href
  {https://ui.adsabs.harvard.edu/abs/2021A&A...656L..15P} {656, L15}

\bibitem[\protect\citeauthoryear{{Planck Collaboration} et~al.,}{{Planck
  Collaboration} et~al.}{2016}]{Planck2016}
{Planck Collaboration} et~al., 2016, \mn@doi [\aap]
  {10.1051/0004-6361/201525830}, \href
  {https://ui.adsabs.harvard.edu/abs/2016A&A...594A..13P} {594, A13}

\bibitem[\protect\citeauthoryear{{Platts}, {Weltman}, {Walters}, {Tendulkar},
  {Gordin}  \& {Kandhai}}{{Platts} et~al.}{2019}]{Platts2019}
{Platts} E.,  {Weltman} A.,  {Walters} A.,  {Tendulkar} S.~P.,  {Gordin}
  J.~E.~B.,   {Kandhai} S.,  2019, \mn@doi [\physrep]
  {10.1016/j.physrep.2019.06.003}, \href
  {https://ui.adsabs.harvard.edu/abs/2019PhR...821....1P} {821, 1}

\bibitem[\protect\citeauthoryear{{Pleunis} et~al.,}{{Pleunis}
  et~al.}{2021}]{Pleunis2021}
{Pleunis} Z.,  et~al., 2021, \mn@doi [\apj] {10.3847/1538-4357/ac33ac}, \href
  {https://ui.adsabs.harvard.edu/abs/2021ApJ...923....1P} {923, 1}

\bibitem[\protect\citeauthoryear{{Prochaska} \& {Zheng}}{{Prochaska} \&
  {Zheng}}{2019}]{Prochaska2019DMhalo}
{Prochaska} J.~X.,  {Zheng} Y.,  2019, \mn@doi [\mnras] {10.1093/mnras/stz261},
  \href {https://ui.adsabs.harvard.edu/abs/2019MNRAS.485..648P} {485, 648}

\bibitem[\protect\citeauthoryear{{Prochaska} et~al.,}{{Prochaska}
  et~al.}{2019}]{Prochaska2019}
{Prochaska} J.~X.,  et~al., 2019, \mn@doi [Science] {10.1126/science.aay0073},
  \href {https://ui.adsabs.harvard.edu/abs/2019Sci...366..231P} {366, 231}

\bibitem[\protect\citeauthoryear{{Ravi}}{{Ravi}}{2019}]{Ravi2019}
{Ravi} V.,  2019, \mn@doi [Nature Astronomy] {10.1038/s41550-019-0831-y}, \href
  {https://ui.adsabs.harvard.edu/abs/2019NatAs...3..928R} {3, 928}

\bibitem[\protect\citeauthoryear{{Schmidt}}{{Schmidt}}{1968}]{Schmidt1968}
{Schmidt} M.,  1968, \mn@doi [\apj] {10.1086/149446}, \href
  {https://ui.adsabs.harvard.edu/abs/1968ApJ...151..393S} {151, 393}

\bibitem[\protect\citeauthoryear{{Scholz} \& {Chime/FRB
  Collaboration}}{{Scholz} \& {Chime/FRB Collaboration}}{2020}]{Scholz2020}
{Scholz} P.,  {Chime/FRB Collaboration} 2020, The Astronomer's Telegram, \href
  {https://ui.adsabs.harvard.edu/abs/2020ATel13681....1S} {13681, 1}

\bibitem[\protect\citeauthoryear{{Shannon} et~al.,}{{Shannon}
  et~al.}{2018}]{Shannon2018}
{Shannon} R.~M.,  et~al., 2018, \mn@doi [\nat] {10.1038/s41586-018-0588-y},
  \href {https://ui.adsabs.harvard.edu/abs/2018Natur.562..386S} {562, 386}

\bibitem[\protect\citeauthoryear{{Spitler} et~al.,}{{Spitler}
  et~al.}{2014}]{Spitler2014}
{Spitler} L.~G.,  et~al., 2014, \mn@doi [\apj] {10.1088/0004-637X/790/2/101},
  \href {https://ui.adsabs.harvard.edu/abs/2014ApJ...790..101S} {790, 101}

\bibitem[\protect\citeauthoryear{{Tendulkar} et~al.,}{{Tendulkar}
  et~al.}{2017}]{Tendulkar2017}
{Tendulkar} S.~P.,  et~al., 2017, \mn@doi [\apjl] {10.3847/2041-8213/834/2/L7},
  \href {https://ui.adsabs.harvard.edu/abs/2017ApJ...834L...7T} {834, L7}

\bibitem[\protect\citeauthoryear{{The CHIME/FRB Collaboration} et~al.,}{{The
  CHIME/FRB Collaboration} et~al.}{2021}]{CHIMEcat2021}
{The CHIME/FRB Collaboration} et~al., 2021, arXiv e-prints, \href
  {https://ui.adsabs.harvard.edu/abs/2021arXiv210604352T} {p. arXiv:2106.04352}

\bibitem[\protect\citeauthoryear{{Xu} et~al.,}{{Xu} et~al.}{2021}]{Xu2021}
{Xu} H.,  et~al., 2021, arXiv e-prints, \href
  {https://ui.adsabs.harvard.edu/abs/2021arXiv211111764X} {p. arXiv:2111.11764}

\bibitem[\protect\citeauthoryear{{Yamasaki} \& {Totani}}{{Yamasaki} \&
  {Totani}}{2020}]{Yamasaki2020}
{Yamasaki} S.,  {Totani} T.,  2020, \mn@doi [\apj] {10.3847/1538-4357/ab58c4},
  \href {https://ui.adsabs.harvard.edu/abs/2020ApJ...888..105Y} {888, 105}

\bibitem[\protect\citeauthoryear{{Yao}, {Manchester}  \& {Wang}}{{Yao}
  et~al.}{2017}]{Yao2017}
{Yao} J.~M.,  {Manchester} R.~N.,   {Wang} N.,  2017, \mn@doi [\apj]
  {10.3847/1538-4357/835/1/29}, \href
  {https://ui.adsabs.harvard.edu/abs/2017ApJ...835...29Y} {835, 29}

\bibitem[\protect\citeauthoryear{{Zhang} \& {Zhang}}{{Zhang} \&
  {Zhang}}{2021}]{Zhang2021}
{Zhang} R.~C.,  {Zhang} B.,  2021, arXiv e-prints, \href
  {https://ui.adsabs.harvard.edu/abs/2021arXiv210907558Z} {p. arXiv:2109.07558}

\bibitem[\protect\citeauthoryear{{Zhang}, {Zhang}, {Li}  \& {Lorimer}}{{Zhang}
  et~al.}{2021}]{Zhangetal2021}
{Zhang} R.~C.,  {Zhang} B.,  {Li} Y.,   {Lorimer} D.~R.,  2021, \mn@doi
  [\mnras] {10.1093/mnras/staa3537}, \href
  {https://ui.adsabs.harvard.edu/abs/2021MNRAS.501..157Z} {501, 157}

\bibitem[\protect\citeauthoryear{{Zhou}, {Li}, {Wang}, {Fan}  \& {Wei}}{{Zhou}
  et~al.}{2014}]{Zhou2014}
{Zhou} B.,  {Li} X.,  {Wang} T.,  {Fan} Y.-Z.,   {Wei} D.-M.,  2014, \mn@doi
  [\prd] {10.1103/PhysRevD.89.107303}, \href
  {https://ui.adsabs.harvard.edu/abs/2014PhRvD..89j7303Z} {89, 107303}

\bibitem[\protect\citeauthoryear{{Zhu}, {Feng}  \& {Zhang}}{{Zhu}
  et~al.}{2018}]{Zhu2018}
{Zhu} W.,  {Feng} L.-L.,   {Zhang} F.,  2018, \mn@doi [\apj]
  {10.3847/1538-4357/aadbb0}, \href
  {https://ui.adsabs.harvard.edu/abs/2018ApJ...865..147Z} {865, 147}

\makeatother
\end{thebibliography}
\input{CHIME_LF_mnras.bbl}

%\begin{thebibliography}{99}
%\end{thebibliography}

%%%%%%%%%%%%%%%%%%%%%%%%%%%%%%%%%%%%%%%%%%%%%%%%%%

%%%%%%%%%%%%%%%%% APPENDICES %%%%%%%%%%%%%%%%%%%%%
\appendix
\section{FRB catalogue}
In this work, we derived model-dependent physical parameters of new CHIME FRBs such as redshift, $E_{\rm rest,400}$, and $\rho_{\rm corr}$.
These parameters are publicly available together with the original observed parameters \citep{CHIMEcat2021} as a single catalogue.
Here we describe the new columns added by this work \citep[see also][for other column names]{CHIMEcat2021}.
\begin{itemize}
\item {\it subw\_upper\_flag}: {\it width\_fitb} and {\it logsubw\_int\_rest} indicate upper limits if 1. Otherwise 0. Different for sub-bursts.
\item {\it scat\_upper\_flag}: {\it scat\_time} indicates the upper limit if 1. Otherwise 0. Common for sub-bursts with the same {\it tns\_name}.
\item {\it spec\_z}: spectroscopic redshift if available. Otherwise $-9999$. Common for sub-bursts with the same {\it tns\_name}.
\item {\it spec\_z\_flag}: {\it spec\_z} is available if 1. Otherwise 0. Common for sub-bursts with the same {\it tns\_name}.
\item {\it E\_obs} (Jy ms Hz): fluence (Jy ms) $\times$ $\Delta\nu$, where $\Delta\nu$ is the band width of 400 MHz at observer's frame. Common for sub-bursts with the same {\it tns\_name}.
\item {\it E\_obs\_error} (Jy ms Hz): error of {\it E\_obs}. Common for sub-bursts with the same {\it tns\_name}.
\item {\it subb\_flag}: 1 if the row belongs to multiple sub-bursts. 0 means FRBs without sub-bursts. Common for sub-bursts with the same {\it tns\_name}.
\item {\it subb\_p\_flag}: {\it subb\_p\_flag}=1 can be used when sub-burst parameters are used. All rows have 1.
\item {\it common\_p\_flag}: {\it common\_p\_flag}=1 can be used when common parameters are used. For each {\it tns\_name}, the first sub-burst indicates 1. Different for sub-bursts.
\item {\it delta\_nuo\_FRB}: (MHz) observed spectral band width. Common for sub-bursts. i.e., {\it high\_freq}$-${\it low\_freq} for FRBs without multiple sub-bursts and max({\it high\_freq})$-$min({\it low\_freq}) for FRBs with multiple sub-bursts. The latter works because there is no frequency gap between sub-bursts in the catalogue.
\item {\it z\_DM}: redshift derived from a dispersion measure. 50 percentile of the PDF. Common for sub-bursts with the same {\it tns\_name}.
\item {\it z\_DM\_error\_p}:  $+1$ $\sigma$ of {\it z\_DM}. 84.135 percentile of the PDF $-$ {\it z\_DM}. Common for sub-bursts with the same {\it tns\_name}.
\item {\it z\_DM\_error\_m}: $-1$ $\sigma$ of {\it z\_DM}. {\it z\_DM} $-$ 15.865 percentile of the PDF. Common for sub-bursts with the same {\it tns\_name}.
\item {\it E\_obs\_400} (Jy ms Hz): observed energy integrated over 400 MHz at emitter's frame. 50 percentile of the probability distribution function (PDF). Common for sub-bursts with the same {\it tns\_name}.
\item {\it E\_obs\_400\_error\_p} (Jy ms Hz): $+1$ $\sigma$ of {\it E\_obs\_400}.  84.135 percentile of the PDF $-$ {\it E\_obs\_400}. Common for sub-bursts with the same {\it tns\_name}.
\item {\it E\_obs\_400\_error\_m} (Jy ms Hz): $-1$ $\sigma$ of {\it E\_obs\_400}. {\it E\_obs\_400} $-$ 15.865 percentile of the PDF. Common for sub-bursts with the same {\it tns\_name}.
\item {\it logsubw\_int\_rest}: (ms) log rest-frame intrinsic duration of a sub-burst in the logarithmic scale. 50 percentile of the PDF. Different for sub-bursts.
\item {\it logsubw\_int\_rest\_error\_p}: (ms) $+1$ $\sigma$ of {\it logsubw\_int\_rest}. 84.135 percentile of the PDF $-$ {\it subw\_int\_rest}. Different for sub-bursts.
\item {\it logsubw\_int\_rest\_error\_m}: (ms) $-1$ $\sigma$ of {\it logsubw\_int\_rest}. {\it logsubw\_int\_rest} $-$ 15.865 percentile of the PDF. Different for sub-bursts.
\item {\it z}: {\it spec\_z} if available, otherwise {\it z\_DM}. Common for sub-bursts with the same {\it tns\_name}.
\item {\it z\_error\_p}: {\it spec\_z} error if {\it spec\_z} is available, otherwise {\it z\_DM\_error\_p}. Common for sub-bursts with the same {\it tns\_name}.
\item {\it z\_error\_m}: {\it spec\_z} error if {\it spec\_z} is available, otherwise {\it z\_DM\_error\_m}. Common for sub-bursts with the same {\it tns\_name}.
\item {\it logE\_rest\_400}: (erg) radio energy in the logarithmic scale integrated over 400 MHz at emitter's frame. 50 percentile of the PDF. Common for sub-bursts with the same {\it tns\_name}.
\item {\it logE\_rest\_400\_error\_p}: (erg) $+1$ $\sigma$ of {\it logE\_rest\_400}. 84.135 percentile of the PDF $-$ {\it logE\_rest\_400}. Common for sub-bursts with the same {\it tns\_name}.
\item {\it logE\_rest\_400\_error\_m}: (erg) $-1$ $\sigma$ of {\it logE\_rest\_400}. {\it logE\_rest\_400} $-$15.865 percentile of the PDF. Common for sub-bursts with the same {\it tns\_name}.
\item {\it logrhoA}: (Gpc$^{-3}$ yr$^{-1}$) the number density in the logarithmic scale derived by the $V_{\rm max}$ method and {\it redshift bin A}. Uncorrected for the selection functions. Common for sub-bursts with the same {\it tns\_name}.
\item {\it logrhoA\_error\_p}: (Gpc$^{-3}$ yr$^{-1}$) $+1$ $\sigma$ of {\it logrhoA}. 84.135 percentile of the PDF $-$ {\it logrhoA}. Common for sub-bursts with the same {\it tns\_name}.
\item {\it logrhoA\_error\_m}: (Gpc$^{-3}$ yr$^{-1}$) $-1$ $\sigma$ of {\it logrhoA}. {\it logrhoA} $-$ 15.865 percentile of the PDF. Common for sub-bursts with the same {\it tns\_name}.
\item {\it logrhoB}: (Gpc$^{-3}$ yr$^{-1}$) the number density in the logarithmic scale derived by the $V_{\rm max}$ method and {\it redshift bin B}. Uncorrected for the selection functions. Common for sub-bursts with the same {\it tns\_name}.
\item {\it logrhoB\_error\_p}: (Gpc$^{-3}$ yr$^{-1}$) $+1$ $\sigma$ of {\it logrhoB}. 84.135 percentile of the PDF $-$ {\it logrhoB}. Common for sub-bursts with the same {\it tns\_name}.
\item {\it logrhoB\_error\_m}: (Gpc$^{-3}$ yr$^{-1}$) $-1$ $\sigma$ of {\it logrhoB}. {\it logrhoB} $-$ 15.865 percentile of the PDF. Common for sub-bursts with the same {\it tns\_name}.
\item {\it weight\_DM}: weight factor of DM$_{\rm obs}$. Common for sub-bursts with the same {\it tns\_name}.
\item {\it weight\_DM\_error\_p}: $+1$ $\sigma$ of {\it weight\_DM}. 84.135 percentile of the PDF $-$ {\it weight\_DM}. Common for sub-bursts with the same {\it tns\_name}.
\item {\it weight\_DM\_error\_m}: $-1$ $\sigma$ of {\it weight\_DM}. {\it weight\_DM} $-$15.865 percentile of the PDF. Common for sub-bursts with the same {\it tns\_name}.
\item {\it weight\_scat}: weight factor of $\tau_{\rm scat}$. Common for sub-bursts with the same {\it tns\_name}.
\item {\it weight\_scat\_error\_p}: $+1$ $\sigma$ of {\it weight\_scat}. 84.135 percentile of the PDF $-$ {\it weight\_scat}. Common for sub-bursts with the same {\it tns\_name}.
\item {\it weight\_scat\_error\_m}: $-1$ $\sigma$ of {\it weight\_scat}. {\it weight\_scat} $-$15.865 percentile of the PDF. Common for sub-bursts with the same {\it tns\_name}.
\item {\it weight\_w\_int}: weight factor of $w_{\rm int}$. Common for sub-bursts with the same {\it tns\_name}.
\item {\it weight\_w\_int\_error\_p}: $+1$ $\sigma$ of {\it weight\_w\_int}. 84.135 percentile of the PDF $-$ {\it weight\_w\_int}. Common for sub-bursts with the same {\it tns\_name}.
\item {\it weight\_w\_int\_error\_m}: $-1$ $\sigma$ of {\it weight\_w\_int}. {\it weight\_w\_int} $-$15.865 percentile of the PDF. Common for sub-bursts with the same {\it tns\_name}.
\item {\it weight\_fluence}: weight factor of fluence. Common for sub-bursts with the same {\it tns\_name}.
\item {\it weight\_fluence\_error\_p}: $+1$ $\sigma$ of {\it weight\_fluence}. 84.135 percentile of the PDF $-$ {\it weight\_fluence}. Common for sub-bursts with the same {\it tns\_name}.
\item {\it weight\_fluence\_error\_m}: $-1$ $\sigma$ of {\it weight\_fluence}. {\it weight\_fluence} $-$15.865 percentile of the PDF. Common for sub-bursts with the same {\it tns\_name}.
\item {\it weight}: {\it weight\_DM}$\times${\it weight\_scat}$\times${\it weight\_w\_int}$\times${\it weight\_fluence}. Common for sub-bursts with the same {\it tns\_name}.
\item {\it weight\_error\_p}: $+1$ $\sigma$ of {\it weight}. 84.135 percentile of the PDF $-$ {\it weight}. Common for sub-bursts with the same {\it tns\_name}.
\item {\it weight\_error\_m}: $-1$ $\sigma$ of {\it weight}. {\it weight} $-$15.865 percentile of the PDF. Common for sub-bursts with the same {\it tns\_name}.
\item {\it weighted\_logrhoA}: (Gpc$^{-3}$ yr$^{-1}$) the number density in the logarithmic scale ({\it redshift bin A}) corrected for the selection functions. Common for sub-bursts with the same {\it tns\_name}.
\item{\it weighted\_logrhoA\_error\_p}: (Gpc$^{-3}$ yr$^{-1}$) $+1$ $\sigma$ of {\it weighted\_logrhoA}. 84.135 percentile of the PDF $-$ {\it weighted\_logrhoA}. Common for sub-bursts with the same {\it tns\_name}.
\item{\it weighted\_logrhoA\_error\_m}: (Gpc$^{-3}$ yr$^{-1}$) $-1$ $\sigma$ of {\it weighted\_logrhoA}. {\it weighted\_logrhoA} $-$15.865 percentile of the PDF. Common for sub-bursts with the same {\it tns\_name}.
\item {\it weighted\_logrhoB}: (Gpc$^{-3}$ yr$^{-1}$) the number density in the logarithmic scale ({\it redshift bin B}) corrected for the selection functions. Common for sub-bursts with the same {\it tns\_name}.
\item{\it weighted\_logrhoB\_error\_p}: (Gpc$^{-3}$ yr$^{-1}$) $+1$ $\sigma$ of {\it weighted\_logrhoB}. 84.135 percentile of the PDF $-$ {\it weighted\_logrhoB}. Common for sub-bursts with the same {\it tns\_name}.
\item{\it weighted\_logrhoB\_error\_m}: (Gpc$^{-3}$ yr$^{-1}$) $-1$ $\sigma$ of {\it weighted\_logrhoB}. {\it weighted\_logrhoB} $-$15.865 percentile of the PDF. Common for sub-bursts with the same {\it tns\_name}.
\end{itemize}

\section{Possible contamination in the non-repeating FRB sample}
\label{contamination}
A non-repeating FRB is observationally defined by the one-off detection of a radio burst for each FRB source.
This does not necessarily mean either the radio burst happened only one time in the past or another burst will never happen.
In this sense, the non-repeating FRB sources are likely contaminated more or less by repeating FRB sources.
Here we try to reduce the possible contamination of repeating FRB sources in our non-repeating FRB sample.

\citet{Chen2022} applied an unsupervised machine learning classification to the CHIME new FRB catalogue \citep{CHIMEcat2021} using observed parameters such as dispersion measure, fluence, intrinsic duration, scattering, spectral index, spectral running, peak frequency, and minimum/maximum frequencies together with model-dependent parameters derived in this work (e.g. redshift and $E_{\rm rest,400}$).
The Uniform Manifold Approximation and Projection (UMAP) is utilised by \citet{Chen2022}.
They found that the CHIME new FRBs show nine different clustering groups in the projected hyper-dimensions of UMAP to identify three groups including both repeating and (apparently) non-repeating FRBs.
Four isolated groups of non-repeating FRBs \citep[{\it other\_cluster\_1}, {\it other\_cluster\_2}, {\it other\_cluster\_4}, and {\it other\_cluster\_6} in][]{Chen2022} do not include any repeating FRBs, suggesting that they are less contaminated by repeating FRBs.
We use these groups of non-repeating FRBs to derive their energy functions and volumetric rates as a function of redshift.
Because the feature importance of redshift is low in the UMAP classification \citep{Chen2022}, the selection effect on the redshift would be less significant, whereas the spectral shape is the most important factor in classifying the non-repeating and repeating FRBs \citep{Chen2022}.
The analysis described in Section \ref{vmax} is applied to this sample.

\begin{figure}
    \includegraphics[width=\columnwidth]{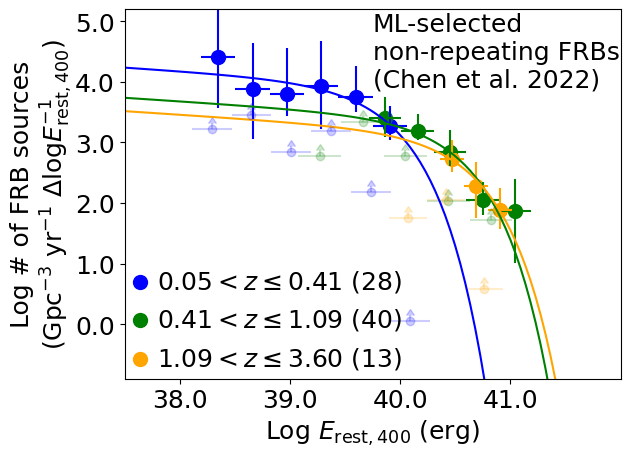}
    \caption{
    Same as Fig. \ref{fig6}(c) except for non-repeating FRBs selected from an unsupervised machine learning (ML) approach by \citet{Chen2022}.
    The selected groups of non-repeating FRBs are clustered in the projected hyper-dimensions of UMAP without any contamination of repeating FRBs \citep{Chen2022}.
    The selected groups include {\it other\_cluster\_1}, {\it other\_cluster\_2}, {\it other\_cluster\_4}, and {\it other\_cluster\_6} \citep{Chen2022}.
    {\it Redshift bin B} is utilised due to the relatively smaller number of sample than that in Section \ref{sample_EF}.
    }
    \label{figB1}
\end{figure}

\begin{figure}
    \includegraphics[width=\columnwidth]{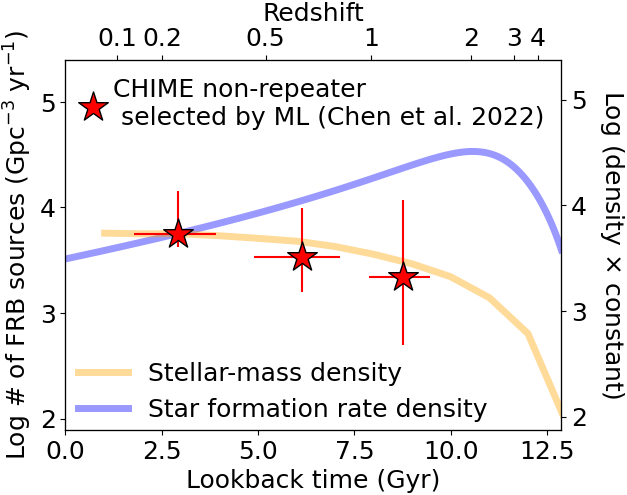}
    \caption{
    Same as Fig. \ref{fig7} except for non-repeating FRBs selected from an unsupervised machine learning (ML) approach by \citet{Chen2022}.
    Note that the cosmic stellar-mass density and the cosmic star-formation rate density are adjusted at $z=0.24$.
    {\it Redshift bin B} is utilised due to the relatively smaller number of sample than that in Section \ref{sample_EF}.
    }
    \label{figB2}
\end{figure}

Fig. \ref{figB1} and \ref{figB2} show the derived energy functions and volumetric rates as a function of redshift.
Both energy functions and volumetric rates indicate the same trend as that found in Figs \ref{fig6} and \ref{fig7} in the sense that the energy functions and volumetric rates decrease towards higher redshifts.
This suggests that the possible contamination from repeating FRBs might not significantly affect the conclusion for non-repeating FRBs described in Sections \ref{volumetric_FRBrate} and \ref{conclusion}.
The best-fit parameters to the energy functions are summarised in Table \ref{tab1}.

\section{Different assumptions on the Milky Way halo and host galaxy components of dispersion measure}
\label{differentDM}
The dispersion measures contributed from the Milky Way dark matter halo (DM$_{\rm halo}$) and an FRB host galaxy (DM$_{\rm host}$) have not been well understood.
DM$_{\rm halo}$ and DM$_{\rm host}$ might be systematically different from those assumed in Section \ref{calc_redshift} \citep[e.g.][]{Dolag2015,Prochaska2019DMhalo,Yamasaki2020, Li2020}.
Here we test how different assumptions on DM$_{\rm halo}$ and DM$_{\rm host}$ affect our results in Section \ref{results} and conclusions in Sections \ref{volumetric_FRBrate} and \ref{conclusion}. 

\begin{figure}
    \includegraphics[width=\columnwidth]{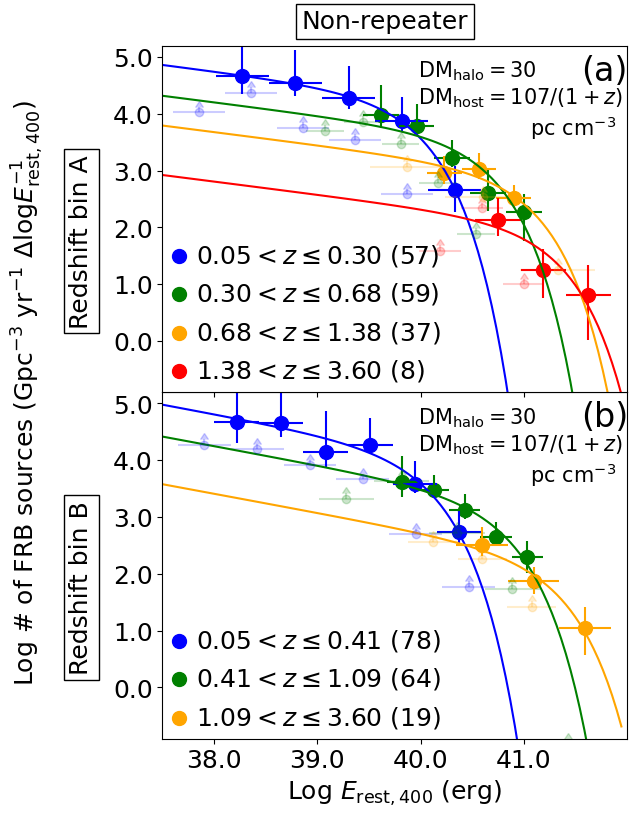}
    \caption{
    Same as Fig. \ref{fig6} (a) and (c) except for the assumptions on DM$_{\rm halo}=30$ pc cm$^{-3}$ \citep{Dolag2015,Prochaska2019DMhalo} and DM$_{\rm host}=107/(1+z)$ pc cm$^{-3}$ \citep{Li2020} to derive the redshifts.
    }
    \label{figC1}
\end{figure}

\begin{figure}
    \includegraphics[width=\columnwidth]{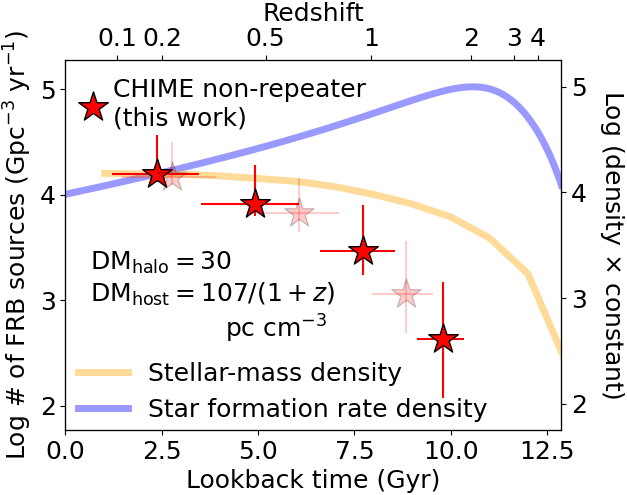}
    \caption{
    Same as Fig. \ref{fig7} except for the assumptions on DM$_{\rm halo}=30$ pc cm$^{-3}$ \citep{Dolag2015,Prochaska2019DMhalo} and DM$_{\rm host}=107/(1+z)$ pc cm$^{-3}$ \citep{Li2020} to derive the redshifts. 
    The cosmic stellar-mass density and the cosmic star-formation rate density are adjusted at $z=0.19$.
    }
    \label{figC2}
\end{figure}

We test DM$_{\rm halo}=30$ pc cm$^{-3}$ \citep{Dolag2015} 
%\textcolor{cyan}{[SY: I suggest to delete \cite{Prochaska2019DMhalo} since DM$_{\rm halo}=30$ pc cm$^{-3}$ is the prediction by \cite{Dolag2015} not by \cite{Prochaska2019DMhalo} (the referee is wrong...). Also, if possible could you please add Yamasaki \& Totani 2020, which is a direction-dependent model calibrated with X-ray observations with its average prediction DM$_{\rm halo}=45$ pc cm$^{-3}$, which is relatively close to \cite{Dolag2015}?]} 
and DM$_{\rm host}=107/(1+z)$ pc cm$^{-3}$ \citep{Li2020} to calculate the energy functions and volumetric non-repeating FRB rates following the analyses described in Section \ref{analysis} and \ref{results}.
The results are shown in Figs. \ref{figC1} and \ref{figC2}.
In Figs. \ref{figC1} and \ref{figC2}, we confirmed decreasing trends of the energy functions and volumetric non-repeating FRB rates towards higher redshifts. 
The $\chi^{2}$ values to the cosmic stellar-mass density are 4.9 and 3.3 for {\it redshift bin A} and {\it redshift bin B}, respectively. 
The corresponding $p$-values are 0.03 and 0.07, indicating that the null hypothesis is not rejected with the 1\% significance threshold. 
The $\chi^{2}$ values to the cosmic star-formation rate density are 21.3 and 14.7 for {\it redshift bin A} and {\it redshift bin B}, respectively. 
The corresponding $p$-values are 3.8e$-6$ and 1.3e$-4$, indicating that the null hypothesis is ruled out with a more than 99\% confidence level.

We also test DM$_{\rm halo}=65$ pc cm$^{-3}$ \citep{Prochaska2019} and DM$_{\rm host}=107/(1+z)$ pc cm$^{-3}$ \citep{Li2020}, confirming almost the same decreasing trend of the volumetric non-repeating FRB rates towards higher redshifts as that presented in Fig. \ref{figC2}.
The calculated $p$-values to the cosmic stellar-mass density (the cosmic star-formation rate density) are 0.02 and 0.06 (3.3e$-6$ and 8.6e$-5$) for {\it redshift bin A} and {\it redshift bin B}, respectively.
Therefore, we conclude that the different assumptions on DM$_{\rm halo}$ and DM$_{\rm host}$ do not significantly affect our conclusions in Sections \ref{volumetric_FRBrate} and \ref{conclusion}.

% Don't change these lines
\bsp	% typesetting comment
\label{lastpage}
\end{document}